\documentclass{emulateapj}
\usepackage{apjfonts}
\slugcomment{accepted for publication in ApJ}
\shorttitle{Statistical Properties of GRB Polarization}
\shortauthors{Toma et al.}
\begin{document}
\title{
Statistical Properties of Gamma-Ray Burst Polarization
}
\author{Kenji Toma\altaffilmark{1,2}, Takanori Sakamoto\altaffilmark{3,4}, 
Bing Zhang\altaffilmark{5}, Joanne E. Hill\altaffilmark{3,6},
Mark L. McConnell\altaffilmark{7}, Peter F. Bloser\altaffilmark{7},
Ryo Yamazaki\altaffilmark{8}, Kunihito Ioka\altaffilmark{9},
and Takashi Nakamura\altaffilmark{10}}

\altaffiltext{1}{Department of Astronomy and Astrophysics, Pennsylvania State 
University, 525 Davey Lab, University Park, PA 16802, USA}
\altaffiltext{2}{Division of Theoretical Astronomy, National Astronomical 
Observatory of Japan (NAOJ), 2-21-1 Osawa, Mitaka, Tokyo 181-8588, Japan}
\altaffiltext{3}{CRESST and NASA Goddard Space Flight Center, Greenbelt, MD 20771, 
USA}
\altaffiltext{4}{Joint Center for Astrophysics, University of Maryland, Baltimore 
County, 1000 Hilltop Circle, Baltimore, MD 21250, USA}
\altaffiltext{5}{Department of Physics and Astronomy, University of Nevada 
Las Vegas, Las Vegas, NV 89154, USA}
\altaffiltext{6}{Universities Space Research Association, 10211 Wincopin Circle, 
Suite 500, Columbia, MD, 21044-3432, USA}
\altaffiltext{7}{Space Science Center, University of New Hampshire, Durham,
NH 03824, USA}
\altaffiltext{8}{Department of Physical Science, Hiroshima University, 
Higashi-Hiroshima, Hiroshima 739-8526, Japan}
\altaffiltext{9}{Theory Division, KEK (High Energy Accelerator Research
Organization), 1-1 Oho, Tsukuba 305-0801, Japan}
\altaffiltext{10}{Department of Physics, Kyoto University, Kyoto 606-8502, Japan}
\email{toma@astro.psu.edu}
\begin{abstract}
The emission mechanism and the origin and structure of magnetic fields in 
gamma-ray burst (GRB) jets are among the most important open questions concerning
the nature of the central engine of GRBs.
In spite of extensive observational efforts, these questions remain to be 
answered and are difficult or even impossible to infer with the spectral and
lightcurve information currently collected.
Polarization measurements will lead to unambiguous answers to several of these questions.
Recent developments in X-ray and $\gamma$-ray polarimetry techniques
have demonstrated a significant increase in sensitivity enabling several 
new mission concepts,
e.g. {\it POET} (Polarimeters for Energetic Transients), providing wide field of 
view and broadband polarimetry measurements.
If launched, missions of this kind would finally
provide definitive measurements of GRB polarizations.
We perform Monte Carlo simulations to derive the distribution of 
GRB polarizations in three emission models;
the synchrotron model with a globally ordered magnetic
field (SO model), the synchrotron model with a 
small-scale random magnetic field
(SR model), and the Compton drag model (CD model).
The results show that {\it POET}, 
or other polarimeters with similar capabilities,
can constrain the GRB emission models by using the statistical properties
of GRB polarizations.
In particular, the ratio of the number of GRBs for which the polarization degrees
can be measured to the number of GRBs that are detected ($N_m/N_d$) and
the distributions of the polarization degrees ($\Pi$) can be used as the criteria.
If $N_m/N_d > 30\%$ and $\Pi$ is clustered between 0.2 and 0.7, the SO model
will be favored.
If instead $N_m/N_d < 15\%$, then the SR or CD model will be favored.
If several events with $\Pi > 0.8$ are observed, then the CD model will be favored.
\end{abstract}


\section{Introduction}
\label{sec:intro}

Gamma-ray bursts (GRBs) are brief, intense flashes of $\gamma$-rays
originating at cosmological distances, and they are the most luminous 
objects in the universe.
They also have broadband afterglows long-lasting after
the $\gamma$-ray radiation has ceased.
It has been established 
that the bursts and afterglows are
emitted from outflows moving towards us 
at highly relativistic speeds \citep{taylor04},
and at least some GRBs are associated with the collapse of
massive stars \citep[e.g.,][]{hjorth03,stanek03}.
Observations suggest that the burst is produced by internal
dissipation within the relativistic jet that is launched from the center
of the explosion, and the afterglow is the synchrotron emission of electrons
accelerated in a collisionless shock driven by the interaction of the jet
with the surrounding medium 
\citep[for recent reviews,][]{piran05,meszaros06,zhang07}.

In spite of extensive observational and theoretical efforts, 
several key questions concerning the nature of the
central engines of the relativistic jets and the jets themselves remain
poorly understood. 
In fact, some of these questions are very difficult or 
even impossible to answer with the spectral and lightcurve information
currently collected. 
On the other hand, polarization information, if retrieved,
would lead to unambiguous answers to these questions. 
In particular, 
polarimetric observations of GRBs can address the following:

{\it Magnetic composition of GRB jets --}
It is highly speculated that strong magnetic
fields are generated at the GRB central engine, and may play an essential
role in the launch of the relativistic jets. 
However, it is unclear whether the burst
emission region is penetrated by a globally structured, dynamically important
magnetic field, and whether the burst is due to shock dissipation or magnetic
reconnection \citep[e.g.,][]{spruit01,zhang02,lyutikov03}.

{\it Emission mechanisms of the bursts --}
The leading model for the emission mechanism of the prompt burst emission is 
synchrotron emission
from relativistic electrons in a globally ordered magnetic field carried from 
the central engine, or random magnetic fields generated in-situ in the shock
dissipation region \citep{rees94}. 
Other suggestions include Compton drag of ambient soft photons
\citep{shaviv95,eichler03,levinson04,lazzati04},
synchrotron self-Compton emission \citep{panaitescu00}, 
and the combination of a thermal
component from the photosphere and a non-thermal component (e.g., synchrotron)
\citep{ryde06,thompson07,ioka07}.

{\it Geometric structure of GRB jets --}
Although it is generally believed that GRB
outflows are collimated, the distribution of the jet
opening angles, the observer's viewing direction, and whether there
are small-scale structures within the global jet are not well understood
\citep{zhang04,yamazaki04,toma05}.

To date, robust positive detections of GRB polarization have been made only
in the optical band in the afterglow phase.
Varying linear polarizations have been observed in several optical afterglows
several hours after the burst trigger, with a level of $\sim 1-3\%$, 
which is consistent with the synchrotron emission mechanism of GRB afterglow
\citep[for reviews, see][]{covino04,lazzati06}.
An upper limit $(< 8\%)$ has been obtained for the early $(t\sim200~{\rm s})$
optical afterglow of GRB 060418 \citep{mundell07}.
Also for radio afterglows, we have several upper limits for the polarization degree
\citep{taylor05,granot05} \citep[for some implications, see][]{toma08}.
As for the prompt burst emission, strong linear polarization of 
the $\gamma$-ray emission
at a level of $\Pi = 80 \pm 20\%$ was claimed for GRB 021206 based
on an analysis of {\it RHESSI} data \citep{coburn03}, 
although this claim remains controversial because of 
large systematic uncertainties \citep{rutledge04,wigger04}.
Several other reports of high levels of polarization in the prompt burst emission are
also statistically inconclusive \citep{willis05,kalemci07,mcglynn07}.

Recently, more sensitive observational techniques for X-ray and $\gamma$-ray 
polarimetry have been developed, and there are several polarimeter mission concepts.
These include Polarimeters for Energetic Transients 
\citep[{\it POET},][]{hill08,bloser08},
Polarimeter of Gamma-ray Observer
\citep[{\it PoGO},][]{mizuno05},
{\it POLAR} \citep{produit05},
Advanced Compton Telescope \citep[{\it ACT},][]{boggs06},
Gravity and Extreme Magnetism \citep[{\it GEMS},][]{jahoda07}  
{\it XPOL} \citep{costa07}, 
Gamma-Ray Burst Investigation via Polarimetry and Spectroscopy
\citep[{\it GRIPS},][]{greiner08},
and so on.

Several of these missions, if launched, would
provide definitive detections of the burst
polarizations and enable us to discuss the statistical properties
of the polarization degrees and polarization spectra. 
Although there are several polarimetry mission concepts described 
in the literature, {\it POET} is the only one to date that incorporates 
a broadband capability for measuring the prompt emission from 
GRBs, and for this reason it provides a good case study for our simulations.
{\it POET} will make measurements with two different polarimeters,
both with wide fields of view.
The Gamma-Ray Polarimeter Experiment (GRAPE; 60-500~keV) and the
Low Energy Polarimeter (LEP; 2-15~keV) provide a broad energy range 
for the observations.
Suborbital versions of both {\it POET} instruments are currently being prepared
for flight within the next few years.  
GRAPE will fly on a sub-orbital balloon in 2011, and the Gamma-Ray Burst
Polarimeter (GRBP, a smaller version of LEP) will fly on a sounding rocket.

Theoretically, it has been shown that 
similarly high levels of linear polarization can
be obtained in several GRB prompt emission models;
the synchrotron model with a globally ordered magnetic field, 
the synchrotron model with a small-scale random magnetic field 
\citep{granot03,lyutikov03,nakar03}, and the Compton drag model
\citep{lazzati04,eichler03,levinson04,shaviv95}.
Thus the detections of GRB prompt emission polarization would 
support these three models.
In this paper, we show that these models can be distinguished
by their statistical properties of observed polarizations.
We performed detailed calculations of the distribution of 
polarization degrees by including realistic spectra of GRB prompt emission
and assuming realistic distributions of the physical parameters of GRB jets,
and show that {\it POET}, or other polarimeters with similar capabilities,
can constrain the GRB emission models.
We use the limits of {\it POET} for GRB detection and polarization measurements
as realistic and fiducial limits.
This paper is organized as follows.
We first introduce the {\it POET} mission concept
in \S~\ref{sec:poet}.
In \S~\ref{sec:theory}, we summarize the properties of the observed
linear polarization from uniform jets within the three emission models.
Based on these models, we perform Monte Carlo simulations of observed
linear polarizations and show how the statistical properties of
observed polarization may constrain GRB emission mechanisms
in \S~\ref{sec:simulation}.
A summary and discussion are given in \S~\ref{sec:summary}.

\section{Properties of POET satellite}
\label{sec:poet}

{\it POET} (Polarimeters for Energetic Transients) is a Small Explorer (SMEX) mission 
concept, that will provide highly sensitive 
polarimetric observations of GRBs and can also make polarimetry measurements of 
solar flares, pulsars, soft gamma-ray repeaters, and slow transients. 
The payload consists of two wide field of view (FoV) instruments: a Low Energy 
Polarimeter (LEP) capable of polarization measurements in the 2-15 keV energy range 
and a high energy polarimeter (Gamma-Ray Polarimeter Experiment; GRAPE) 
that will measure polarization in the 60-500 keV energy range. 
{\it POET} can measure GRB spectra from 2 keV up to 1 MeV. 
The {\it POET} spacecraft provides a zenith-pointed platform for maximizing the 
exposure to deep space and
spacecraft rotation provides a means of effectively dealing with systematics 
in the polarization response. 
{\it POET} provides sufficient sensitivity and sky coverage to detect
up to 200 GRBs in a two-year mission.

LEP and GRAPE determine polarization by measuring the number of events versus 
the event azimuth angle (EAA) as projected onto the sky. 
This is referred to as a modulation profile and represents a measure of the 
polarization magnitude and direction of polarization for the incident beam. 
Depending on the type of polarimeter, the EAA is either the direction of the 
ejected photoelectron (LEP) or the direction of the scattered photon (GRAPE). 
The response of a polarimeter to $100\%$ polarized photons can be quantified 
in terms of the modulation factor, $\mu$,
which is given by:
\begin{equation}
\mu=\frac{C_{\rm max}-C_{\rm min}}{C_{\rm max}+C_{\rm min}}  \label{mod}
\end{equation}
Where $C_{\rm max}$ and $C_{\rm min}$ are the maximum and minimum of the modulation 
profile, respectively. 
The polarization fraction ($\Pi$) of the incident flux is obtained by dividing 
the measured modulation by that expected for $100\%$ polarized flux. 
The polarization angle ($\Phi_0$) corresponds either to the maximum of the 
modulation profile (LEP) or the minimum of the modulation profile (GRAPE). 
To extract these parameters from the data, the modulation histograms are fit 
to the functional form:
\begin{equation}
C(\Phi)=A+B\cos^2(\Phi-\Phi_0)  \label{C_phi}
\end{equation}

The sensitivity of a polarimeter is defined in terms of the 
minimum detectable polarization (MDP), which refers 
to the minimum level of polarization that is detectable with a given 
observation (or, equivalently, the apparent polarization arising from 
statistical fluctuations in unpolarized data). 
The precise value of the MDP will depend on the source parameters (fluence, 
spectrum, etc.) and the polarimeter characteristics. 
At the $99\%$ confidence level, the
MDP can be expressed as,
\begin{equation}
MDP = \frac{4.29}{\mu R_s} \sqrt{\frac{R_s + R_b}{t}},
\label{eq:MDP}
\end{equation}
where $R_s$ is the observed source strength (${\rm cts}~{\rm s}^{-1}$), 
$R_b$ is the total observed background rate (${\rm cts}~{\rm s}^{-1}$), and
$t$ is the observing time (s).
The ultimate sensitivity, however, may not be limited by statistics but by 
systematic errors created by false modulations that arise from azimuthal 
asymmetries in the instrument.

\begin{deluxetable}{lcc}
\tablewidth{0pt}
\tablecaption{
Instrument Parameters
\label{tabparam}
}
\tabletypesize{\scriptsize}
\tablehead{
  \colhead{} &
  \colhead{GRAPE} &
  \colhead{LEP} \\
}
\startdata
Polarimetry    & 60--500 keV & 2--15 keV\\
Detectors & BGO/plastic scintillator (62) & $Ne:CO_2:CH_3NO_2$ Gas (8)\\
Spectroscopy &  15 keV -- 1 MeV & 2 -- 15 keV\\
Detectors & NaI(TI) scintillator (2) & as above\\
Field-of-View & $\pm60^o$ & $\pm44^o$  \\
\enddata
\end{deluxetable}

At energies from $\sim50$ keV up to several MeV, photon interactions are 
dominated by Compton scattering. 
The operational concept for GRAPE is based on the fact that, in Compton 
scattering, photons are preferentially scattered at a right angle to 
the incident electric field vector (the polarization vector)
\citep{bloser08,bloser06,jason05}. 
If the incident beam of photons is polarized, the azimuthal distribution 
of scattered photons will be asymmetric.  
The direction of the polarization vector is defined by the minimum of 
the scatter angle distribution. 
The GRAPE performance characteristics are shown in Table \ref{tabparam}.
The design of the GRAPE instrument is very modular, with 62 independent 
polarimeter modules and 2 spectroscopy modules. 
Each polarimeter module incorporates an array of optically independent 
5x5x50 $mm^3$ non-hygroscopic scintillator elements aligned with and 
optically coupled to the 8x8 scintillation light sensors of a 64-channel 
MAPMT. 
Two types of scintillators are employed. Low-Z plastic scintillator is 
used as an effective medium for Compton scattering. 
High-Z inorganic scintillator (Bismuth Germanate, BGO) is used as a 
calorimeter, for absorbing the full energy of the scattered photon. 
The arrangement of scintillator elements within a module has 28 BGO 
calorimeter elements surrounding 32 plastic scintillator scattering 
elements. 
Valid polarimeter events are those in which a photon Compton scatters 
in one of the plastic elements and is subsequently absorbed in one of 
the BGO elements. 
These events can be identified as a coincident detection between one 
plastic scintillator element and one BGO calorimeter element. 
The azimuthal scatter angle is determined for each valid event by 
the relative locations of hit scintillator elements. 
It is not necessary to know where within each element the interaction
takes place (e.g., the depth of interaction). 
It is sufficient to know only the lateral location of each element to 
generate a histogram of photon scatter angles.

At energies below $\sim 50$~keV, the most sensitive technique
for broadband polarimetry is the photoelectric effect.
The LEP measures the polarization of incident photons with the 
innovative operation of a Time Projection Chamber (TPC) 
\citep{black07}. 
The LEP polarimeter enclosure consists of four dual-readout detector 
modules each with an isolated gas volume contained by a Be X-ray window. 
Each detector module contains two 6 x 12 x 24 $cm^3$ (LxWxH) TPCs that 
share a single X-ray transparent drift electrode. 
Each TPC is comprised of a micropattern proportional counter, 
consisting of a shared drift electrode and a high-field gas electron 
multiplier (GEM) positioned 1 mm from a strip readout plane. 
When an X-ray is absorbed in the gas between the drift electrode and 
the GEM, a photoelectron is ejected in a preferential direction with 
a $\cos^2\Phi$ distribution, where $\Phi$ is the azimuthal angle 
measured from the X-ray polarization vector. 
As the photoelectron travels through the gas it creates a path of 
ionization that drifts in a moderate, uniform field to the GEM where 
an avalanche occurs. 
The charge finally drifts to the strip detector where it is read out.

\begin{figure}
\epsscale{1.2}
\plotone{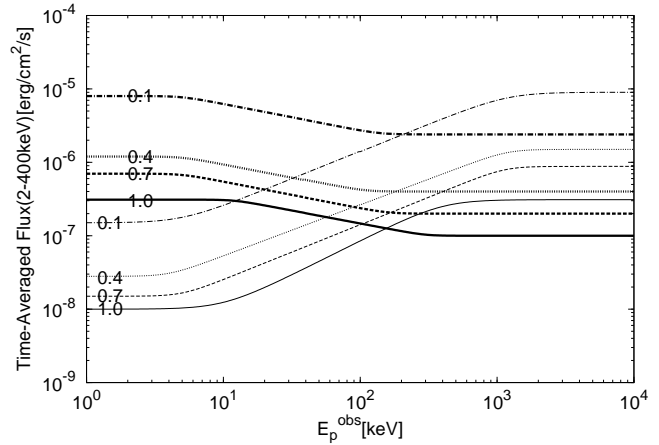}
\caption{
Contour of MDPs (i.e., the minimum values of polarization that is detectable 
at the 99\% confidence level)
of GRAPE ({\it thick lines}) and LEP ({\it thin lines})
for variable $E_{p,{\rm obs}}$ and $F$ and for fixed
$\alpha=-0.2, \beta=1.2, T=20$~sec and the incident angle $30$~degree,
where $E_{p,{\rm obs}}$ is the observed photon energy at the spectral peak,
$F$ is the time-averaged flux in 2-400 keV, $\alpha$ and $\beta$ are the
lower and higher indices of the $F_{\nu}$ spectrum, respectively, and
$T$ is the duration of the prompt emission. 
The combination of GRAPE and LEP enable us to measure polarizations
with reasonable sensitivity in very wide energy range.
}
\label{fig:mdp}
\end{figure}

To estimate realistic MDP values for GRBs detected by GRAPE and LEP, we perform an 
analytical calculation for LEP and a Monte Carlo simulation for GRAPE using the 
current instrument configuration (Table \ref{tabparam}). 
The input spectrum in the calculation and the simulation is a typical GRB 
spectrum which can be described as a smoothly broken power-law spectra 
characterized by photon energy at the $\nu F_{\nu}$ spectral peak, $E_{p,{\rm obs}}$, 
and lower and higher indices of the $F_{\nu}$ spectrum, $\alpha$ and $\beta$,
respectively \citep{band93}.
(We treat the spectral indices of the specific energy flux $F_{\nu}$, while
\citet{band93} define $\alpha_B$ and $\beta_B$ as the indices of the photon
number flux, i.e., $\alpha = -(\alpha_B+1)$ and $\beta = -(\beta_B+1)$, since
we will calculate the net polarizations by using specific energy fluxes
(equation (\ref{eq:band_spec})).)
The various $E_{p,{\rm obs}}$ and time-averaged flux in 2-400 keV, $F$, are 
investigated with fixed $\alpha = -0.2, \beta=1.2$, and a burst duration of $T=20$ s. 
We also assume the incident angle of bursts to be 30 degree off-axis. 
We interpret simulated events with $\Pi > MDP$ as '$\Pi$-measurable events'. 
Figure 1 shows the contour of the MDP values in the $E_{p,{\rm obs}} - F$ plane for 
GRAPE and LEP. 
As can be seen in the figure, with the combination of LEP and GRAPE, it is possible 
to measure the polarization of GRBs with $E_{p,obs}$ ranging from a few keV to MeV 
with reasonable sensitivity.

\section{Theoretical Models}
\label{sec:theory}

We calculate the linear polarization
for instantaneous emission from a thin spherical shell moving
radially outward with a bulk Lorentz factor $\gamma \gg 1$ and 
an opening angle $\theta_j$.
The comoving-frame emissivity has the functional form of 
$j'^I_{\nu'} = A_0 f(\nu') \delta(t'-t'_0) \delta(r'-r'_0)$,
where $A_0$ is the normalization which may depend on direction in the
comoving frame and other physical quantities of the shell
and $f(\nu')$ represents the spectral shape.
A prime represents the physical quantities in the comoving frame.
The delta functions describe the instantaneous emission at $t=t_0$
and $r=r_0$.
The normalization, $A_0$, has units of ${\rm erg}~{\rm cm}^{-2}~
{\rm str}^{-1}~{\rm Hz}^{-1}$.
Using the spherical coordinate system $(r, \theta, \phi)$ in the lab frame,
where $\theta=0$ is the line of sight, we obtain the
spectral fluence \citep{granot99,woods99,ioka01}:
\begin{equation}
I_{\nu} = \frac{1+z}{d_L^2} \int d\phi \int d(\cos\theta) r_0^2
\frac{A_0 f(\nu')}{\gamma^2(1-\beta\cos\theta)^2},
\label{eq:flu}
\end{equation}
where $z$ and $d_L$ are the redshift and the luminosity distance of the
source, respectively, and $\nu'=(1+z)\nu\gamma(1-\beta\cos\theta)$.
The integration is performed within the jet cone, so that it depends on
the viewing angle $\theta_v$, i.e., 
the angle between the jet axis and the line of sight. 
The corresponding Stokes parameters of the local emission 
(i.e., the emission from a given point on the shell)
are given by $j'^Q_{\nu'} = j'^I_{\nu'} \Pi'_0 \cos(2\chi')$ and 
$j'^U_{\nu'} = j'^I_{\nu'} \Pi'_0 \sin(2\chi')$, where 
$\Pi'_0$ and $\chi'$ are the polarization degree and position angle of
the local emission measured in the comoving frame, respectively.
The Stokes parameters of the emission from the whole shell can be
obtained by integrating those of the local emission similarly to the
intensity $I_{\nu}$:
\begin{equation}
\left\{
\begin{array}{c} Q_{\nu} \\ U_{\nu} 
\end{array}
\right\}
= \frac{1+z}{d_L^2} \int d\phi \int d(\cos\theta) r_0^2
\frac{A_0 f(\nu')}{\gamma^2(1-\beta\cos\theta)^2}
\Pi_0
\left\{
\begin{array}{c} \cos(2\chi) \\ \sin(2\chi)
\end{array}
\right\}.
\end{equation}
The polarization degree is Lorentz invariant, i.e., $\Pi'_0 = \Pi_0$.
The position angle $\chi$ is calculated by taking account of
the Lorentz transformation of the electromagnetic waves, and
it is measured from a fixed direction, which we choose to be the direction
from the line of sight to the jet axis.
Then by calculating $\{I, Q, U\} = \int^{\nu_2}_{\nu_1} d\nu 
\{I_\nu, Q_\nu, U_\nu \}$, we obtain the time-averaged linear polarization
in the given wavebands $[\nu_1, \nu_2]$:
\begin{equation}
\Pi = \frac{\sqrt{Q^2 + U^2}}{I}.
\end{equation}

We consider synchrotron and Compton drag (CD) mechanisms for the GRB prompt 
emission.
In the synchrotron case, the magnetic field consists of a globally ordered 
field, ${\mathbf B}_{\rm ord}$, and small-scale random field, 
${\mathbf B}_{\rm rnd}$, i.e., 
${\mathbf B} = {\mathbf B}_{\rm ord} + {\mathbf B}_{\rm rnd}$.
The field ${\mathbf B}_{\rm ord}$ may originate from the central engine,
while ${\mathbf B}_{\rm rnd}$ may be produced in the emission region itself.
Here we consider two extreme cases; synchrotron model with an ordered field
(SO), in which $B^2_{\rm ord} \gg \langle B^2_{\rm rnd} \rangle$,
and a synchrotron model with a random field (SR), in which 
$B^2_{\rm ord} \ll \langle B^2_{\rm rnd} \rangle$.
For the SO model, in particular, we assume a toroidal magnetic field.
In the following sub-sections, we describe $A_0$, $f(\nu')$, $\Pi_0$, and $\chi$
as functions of $(\theta, \phi)$ for each model, and calculate the 
linear polarization for given parameters $\gamma$, $\theta_j$, 
$\theta_v$, and $z$.

\subsection{SO model: synchrotron with ordered field}
\label{subsec:so}

The prompt emission of GRBs could be explained by synchrotron emission from 
accelerated electrons 
that have a non-thermal energy spectra by some 
dissipation process within the jet, e.g, internal shocks.
Synchrotron emission from the relativistically moving shell within a
globally ordered magnetic field results in a net observed linear polarization,
reflecting the direction of the field
\citep{lyutikov03,granot03,nakar03}.
Let us assume that the jet is permeated by a toroidal field.
This is a likely configuration if a magnetic field is advected 
by the jet with a constant speed from the central engine
\citep[e.g.,][]{spruit01,fendt04}.

A general formula for calculating the observed linear polarization for
synchrotron emission from a uniform jet, in which the electrons have a single
power-law energy spectrum and an isotropic pitch angle distribution and
the magnetic field is ordered globally, is
derived by \citet{granot03} and \citet{granot05}.
Here we adopt their formulation and extend it for the electrons having a broken
power-law energy spectrum in order to reproduce the typical observed spectra of GRBs
\citep{band93}.
We adopt the following form for the radiation spectrum:
$f(\nu')=\tilde{f}(x)$ where $x=\nu'/\nu'_0$ and 
\begin{equation}
\tilde{f}(x) = 
\left\{
\begin{array}{lr}
x^{-\alpha} e^{-x} & {\rm for}~ x \leq \beta-\alpha 
\\
x^{-\beta} (\beta-\alpha)^{\beta-\alpha} e^{\alpha-\beta} 
& {\rm for}~ x \geq \beta-\alpha.
\end{array}
\right.
\label{eq:band_spec}
\end{equation}
where $\nu'_0$, $\alpha$, and $\beta$ are the break frequency and
low-energy and high-energy spectral indices of the comoving spectrum, respectively.
\footnote{
In our model the radiation spectrum (\ref{eq:band_spec}) is thought to be
produced by the broken power-law energy spectrum of electrons:
$N(\gamma_e) \propto \gamma_e^{-p_1}$ for $\gamma_e < \gamma_0$ and 
$N(\gamma_e) \propto \gamma_e^{-p_2}$ for $\gamma_e > \gamma_0$, where
$\alpha = (p_1-1)/2$ and $\beta = (p_2-1)/2$.
This formulation also includes the case of $p_1<1/3$, in which
$\alpha=-1/3$, $A_0 \propto (\sin\theta'_B)^{2/3}$,
and $\Pi_0^{\rm syn} = 1/2$ for $x\leq \beta-\alpha$ \citep{granot03}.
}
If we assume that the energy spectrum of the electrons and the 
strength of the magnetic field are uniform in the emitting shell,
then we may write $A_0 = (\sin\theta'_B)^{\alpha+1}$, where
$\theta'_B$ is the angle between the direction of the emitted radiation
and the local direction of the magnetic field \citep{rybicki79}.
The local polarization degree is given by:
\begin{equation}
\Pi_0 = \Pi^{\rm syn}_0 \equiv \left\{
\begin{array}{lr}
(\alpha+1)/(\alpha+\frac{5}{3}) & {\rm for}~x \leq \beta-\alpha \\
(\beta+1)/(\beta+\frac{5}{3}) & {\rm for}~x \geq \beta-\alpha.
\end{array}
\right.
\label{eq:synch_pi}
\end{equation}
For a globally ordered magnetic field, the Faraday depolarization effect may be
strong within the emitting region \citep[e.g.,][]{toma08,matsumiya03,sagiv04}, 
but we neglect it here for simplicity.
By using a new variable $y \equiv (\gamma \theta)^2$, we obtain 
(see Appendix~\ref{subsec:app_so}):
\begin{equation}
\sin\theta'_B = \left[ \left(\frac{1-y}{1+y}\right)^2
+ \frac{4y}{(1+y)^2} \frac{(a-\cos\phi)^2}{1+a^2-2a\cos\phi}\right]^{1/2},
\label{eq:tsinthetab}
\end{equation}
\begin{equation}
\chi = \phi + \arctan \left(\frac{1-y}{1+y}\frac{\sin\phi}{a-\cos\phi}\right),
\label{eq:tchi}
\end{equation}
where $a = \theta/\theta_v$.
Then the formulation of the net polarization degree in the observed
frequency region $[\nu_1, \nu_2]$ becomes:
\begin{equation}
\begin{array}{l}
\Pi = \left|
\int^{\nu_2}_{\nu_1} d\nu
\int^{(1+q)^2 y_j}_0 \frac{dy}{(1+y)^2} 
\right. \\ \left. \times
\int^{\Delta\phi(y)}_{-\Delta\phi(y)} 
d\phi ~\tilde{f}(x) (\sin\theta'_B)^{\alpha+1} \Pi^{\rm syn}_0(x) \cos(2\chi) \right| 
\\
\times \left[
\int^{\nu_2}_{\nu_1} d\nu
\int^{(1+q)^2 y_j}_0 \frac{dy}{(1+y)^2} \int^{\Delta\phi(y)}_{-\Delta\phi(y)}
d\phi ~\tilde{f}(x) (\sin\theta'_B)^{\alpha+1} \right]^{-1},
\end{array}
\label{eq:atorob}
\end{equation}
where  
$x = (1+z)\nu (1+y)/2\gamma \nu'_0$, and
\begin{equation}
q = \frac{\theta_v}{\theta_j}, ~~~
y_j = (\gamma\theta_j)^2,
\end{equation}
\begin{equation}
\Delta\phi(y) = \left\{
\begin{array}{lr}
0 & {\rm for}~q > 1 ~{\rm and}~ y<(1-q)^2 y_j, \\
\pi & {\rm for}~q < 1 ~{\rm and}~ y<(1-q)^2 y_j, \\
\cos^{-1}\left[\frac{(q^2 - 1)y_j + y}{2q\sqrt{y_j y}}\right] & {\rm otherwise}.
\end{array}
\right.
\label{eq:delta_phi}
\end{equation}
The polarization degree, $\Pi$, in the waveband [$\nu_1, \nu_2$] can be calculated
if the geometrical parameters, $y_j, q$, the spectral parameters,
$\gamma\nu'_0, \alpha, \beta$, and the redshift, $z$, are given.

\begin{figure}
\epsscale{1.2}
\plotone{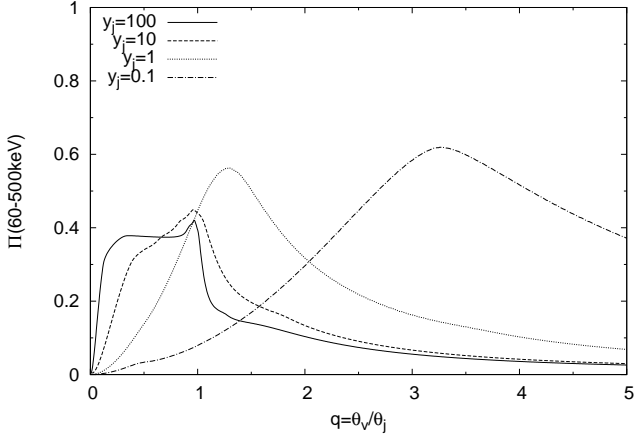}
\caption{
Linear polarization degrees in the $60-500$~keV band 
as a function of $q=\theta_v/\theta_j$,
where $\theta_v$ is the viewing angle of the observer and
$\theta_j$ is the jet opening angle,
for several values of 
$y_j = (\gamma \theta_j)^2$, 
calculated in the SO model (synchrotron model with globally ordered 
magnetic field).
The other parameters are $\gamma \nu'_0 = 350$~keV, $\alpha=-0.2, \beta=1.2$,
and $z=1$.
}
\label{fig:atorob}
\end{figure}

Figure~\ref{fig:atorob} shows the polarization degree in the
$60-500$~keV band as a function of $q$ for several values of $y_j$.
The other parameters are $\gamma \nu'_0 = 350$~keV, $\alpha = -0.2$, 
$\beta=1.2$, and $z=1$.
The polarization degree is negligible for $q \approx 0$, 
because in this case the local polarization vectors are axisymmetric around 
the line of sight, i.e., $\chi = \phi$ (see Appendix~\ref{subsec:app_so}),
and the local polarizations are canceled out.
For $y_j > 1$, a high level of polarization is obtained for 
$y_j^{-1/2} < q < 1$ (i.e., $\gamma^{-1} < \theta_v < \theta_j$).
In this case, only a fraction of the emitting shell (i.e., $\theta < \gamma^{-1}$)
is bright because of the relativistic beaming effect,
and the direction of the magnetic field
is quite ordered in the bright region.
The contribution of the emission from
high latitude, $\theta > \gamma^{-1}$, is negligible, especially for $y_j \geq 100$, 
so that the net polarization
degree is determined only by the emission from the bright region with
$\theta < \gamma^{-1}$ and then it is nearly constant.
The results of our calculations
for the case of $\alpha=\beta$ and $y_j \geq 100$ are consistent
with the results of \citet{granot03} and \citet{lyutikov03}.
For $y_j < 1$, a high level of polarization is obtained for
$q \sim 1+y_j^{-1/2}$ (i.e., $\theta_v \sim \theta_j + \gamma^{-1}$).
In this case, the bright region on the emitting shell is small, also.

The polarization is higher for softer spectra (i.e., larger $\alpha$ and $\beta$).
For example,
for $y_j = 100$, $\gamma \nu'_0 = 350$~keV, and $z=1$,
the polarization degree at the plateau for $q<1$ is
$\simeq 0.28$ for $\alpha=-0.5$ and $\beta = 0.9$, 
while it is $\simeq 0.52$ for $\alpha=0.4$ and $\beta=1.8$.
This is caused mainly by the dependence of the synchrotron polarization on
the spectral indices (equation \ref{eq:synch_pi}).
The maximum polarization degree obtained in the SO model is $\simeq 0.8$
for $y_j \geq 0.01$, $\alpha \leq 0.4$, and $\beta \leq 1.8.$

\subsection{SR model: synchrotron with random field}
\label{subsec:sr}

If the magnetic field is produced at the shock itself within the jet, 
the directions of the field would be random on a scale
as small as the plasma skin depth \citep{gruzinov99,medvedev99}.
It is quite plausible that the directions of the field are not completely
random, but have symmetry around the direction normal to the shock.
The less isotropic the magnetic field directions behind the shock,
the higher the local polarization.
We consider the extreme case in which the field is random strictly
within the plane of the shock.
In this model, the directions of the local polarization vectors on the shell
are axisymmetric around the line of sight (see below),
so that no net polarization remains
if the visible region, $\theta < \gamma^{-1}$, is wholly within the jet cone.
However, if the observer views the jet from an off-axis angle and the symmetry is
broken a high level of polarization remains 
\citep{waxman03,sari99,ghisellini99}.

Similarly to the SO model, we adopt the broken power-law form of the spectrum:
$f(\nu') = \tilde{f}(x),$ where $x = \nu'/\nu'_0$ and $\tilde{f}(x)$ is given
by equation (\ref{eq:band_spec}).
We assume that the energy distribution of 
the electrons and the strength of the magnetic field are uniform
in the emitting shell.
The local Stokes parameters are given by averaging them with respect to 
the magnetic field directions within the shock plane 
(see Appendix~\ref{subsec:app_sr}).
Thus we may write $A_0 = \langle (\sin\theta'_B)^{\alpha+1} \rangle$,
where $\langle \rangle$ represents the average.
The local polarization degree is given by
$\Pi_0 = \Pi_0^{\rm syn} \langle (\sin\theta'_B)^{\alpha+1} \cos(2\phi'_B) \rangle
/\langle (\sin\theta'_B)^{\alpha+1}\rangle$,
where: 
\begin{equation}
\langle (\sin\theta'_B)^{\alpha+1} \rangle = \frac{1}{\pi} \int^\pi_0 d\eta'~
\left[1- \frac{4y}{(1+y)^2} \cos^2\eta'
\right]^{(\alpha+1)/2},
\end{equation}
\begin{equation}
\begin{array}{r}
\langle (\sin\theta'_B)^{\alpha+1} \cos(2\phi'_B) \rangle = \frac{1}{\pi} 
\int^\pi_0 d\eta'~
\left\{
\left[1- \frac{4y}{(1+y)^2} \cos^2\eta'
\right]^{(\alpha-1)/2} 
\right. \\ \times \left.
\left[\sin^2\eta' - \left(\frac{1-y}{1+y}\right)^2 \cos^2\eta' \right] 
\right\}.
\end{array}
\end{equation}
The local polarization position angle measured in the lab frame is given by $\chi = \phi$,
therefore we obtain the formulation for the net polarization in the observed 
frequency region $[\nu_1, \nu_2]$:
\begin{equation}
\begin{array}{l}
\Pi = \left|
\int^{\nu_2}_{\nu_1} d\nu 
\int^{(1+q)^2 y_j}_0 \frac{dy}{(1+y)^2}
\tilde{f}(x) \Pi_0^{\rm syn}(x) 
\right. \\ \left. \times
\langle (\sin\theta'_B)^{\alpha+1} \cos(2\phi'_B) \rangle 
\sin(2\Delta\phi(y)) 
\right| \\
\times \left[
\int^{\nu_2}_{\nu_1} d\nu
\int^{(1+q)^2 y_j}_0 \frac{dy}{(1+y)^2}
\tilde{f}(x) \langle (\sin\theta'_B)^{\alpha+1} \rangle~2\Delta\phi(y)
\right]^{-1},
\end{array}
\label{eq:arndb}
\end{equation}
where $q=\theta_v/\theta_j$, $y_j=(\gamma \theta_j)^2$, 
$x=(1+z)\nu(1+y)/2\gamma\nu'_0$,
and $\Pi_0^{\rm syn}$ and $\Delta\phi(y)$ are given by
equations (\ref{eq:synch_pi}) and (\ref{eq:delta_phi}), respectively.

\begin{figure}
\epsscale{1.2}
\plotone{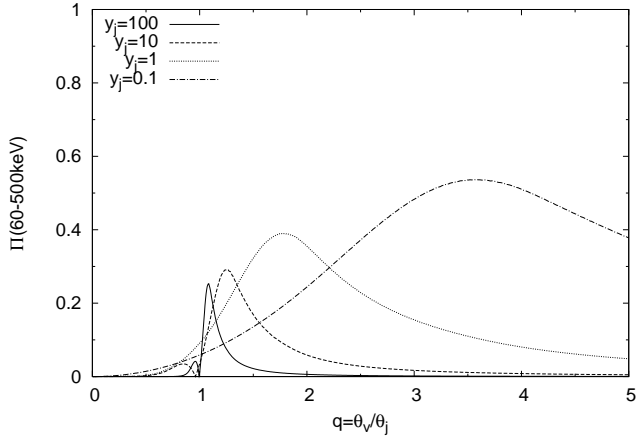}
\caption{
Same as Figure~\ref{fig:atorob}, but in the SR model
(synchrotron model with small-scale random magnetic field).
}
\label{fig:arndb}
\end{figure}

Figure~\ref{fig:arndb} shows the polarization degree in the $60-500$~keV
band as a function of $q$ for several values of $y_j$.
The other parameters are $\gamma\nu'_0 = 350$~keV, $\alpha=-0.2, \beta=1.2$,
and $z=1$.
The results of our calculations for the case of $\alpha=\beta$ 
are consistent with those of \citet{granot03} and \citet{nakar03}.
A high level of polarization is obtained for $q \sim 1+y_j^{-1/2}$ 
(i.e., $\theta_v \sim \theta_j + \gamma^{-1}$) for each value of $y_j$.
Since the local polarization vectors are axisymmetric around the line
of sight, the local polarizations are canceled out if the line of sight
is within the jet cone.
If the jet is observed from an off-axis angle, the net polarization remains.
The local polarization degree is highest for emission where $\theta = \gamma^{-1}$,
so that the net polarization has a maximum value.
The maximum $\Pi$ is higher for smaller $y_j$, because the contribution 
of the emission from high latitude points ($\theta > \gamma^{-1}$),
with a low level of local polarization, is smaller.

Similarly to the SO model, the polarization is higher for softer spectra,
mainly because of the dependence of the local polarization degree on frequency
(equation \ref{eq:synch_pi}).
For example, for $y_j=1$, $\gamma \nu'_0=350$~keV, and $z=1$,
the maximum polarization is 
$\simeq 0.32$ for $\alpha = -0.5$ and $\beta=0.9$, 
while it is $\simeq 0.49$ for $\alpha=0.4$ and $\beta=1.8$.
For $y_j \geq 0.01$, $\alpha \leq 0.4$, and $\beta \leq 1.8,$
the maximum polarization degree in the SR model is $\simeq 0.8$.

\subsection{CD model: Compton drag model}
\label{subsec:compton}

The prompt emission from GRBs could be produced by bulk inverse Comptonization of 
soft photons
from the relativistic jet \citep{lazzati04,eichler03,levinson04,shaviv95}.
The local polarization position angles are symmetric around the line of
sight, similarly to the SR model.
Therefore this model also
requires an off-axis observation of the jet to
achieve a high level of polarization.
However, the CD model is different from the SR model in the fact that
the CD model can in principle achieve $\Pi \sim 1$ under the
most optimistic geometric configurations, whereas the maximum $\Pi$
is $\sim (\beta+1)/(\beta+\frac{5}{3}) \sim 0.8$ in the SR model.

We assume that the seed radiation is unpolarized and has a non-thermal,
isotropic spectrum, and the scattered radiation has the broken power-law spectrum
$f(\nu') = \tilde{f}(x)$, where $x=\nu'/\nu'_0$ and $\tilde{f}(x)$ is
given by equation (\ref{eq:band_spec}).
If the intensity of the seed radiation and the electron number density of the shell
are assumed to be uniform then we may write $A_0 = (1+\cos^2 \theta')/2$, and
$\Pi_0 = (1-\cos^2 \theta')/(1+\cos^2 \theta')$
\citep{rybicki79,begelman87}.
The polarization vectors in the comoving frame are perpendicular to both
incident and scattering directions of photons, so that we obtain
$\chi = \phi + \frac{\pi}{2}$ in the lab frame.
Therefore we achieve the formulation for the net linear polarization
in the observed frequency region $[\nu_1, \nu_2]$:
\begin{equation}
\begin{array}{l}
\Pi = \left|
\int^{\nu_2}_{\nu_1} d\nu
\int^{(1+q)^2 y_j}_0 \frac{dy}{(1+y)^2}
\tilde{f}(x) \frac{2y}{(1+y)^2}
\sin(2\Delta\phi(y)) 
\right| \\
\times \left[
\int^{\nu_2}_{\nu_1} d\nu
\int^{(1+q)^2 y_j}_0 \frac{dy}{(1+y)^2}
\tilde{f}(x) \frac{1+y^2}{(1+y)^2} 2\Delta\phi(y)
\right]^{-1},
\end{array}
\label{eq:acomp}
\end{equation}
where $q=\theta_v/\theta_j$, $y_j=(\gamma \theta_j)^2$, 
$x=(1+z)\nu(1+y)/2\gamma\nu'_0$, and 
$\Delta\phi(y)$ is given by equation (\ref{eq:delta_phi}).

\begin{figure}
\epsscale{1.2}
\plotone{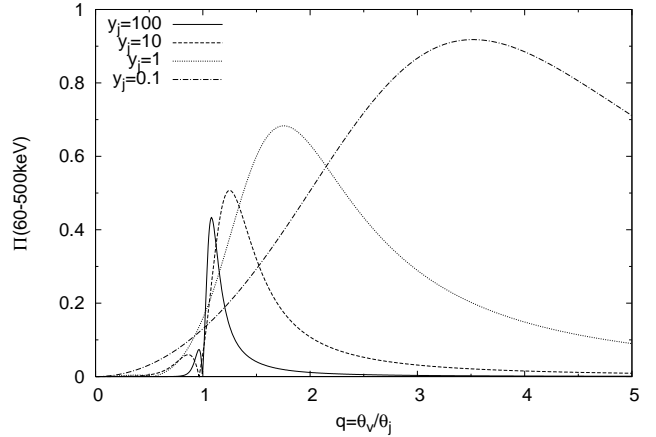}
\caption{
Same as Figure~\ref{fig:atorob}, but in the CD model
(Compton drag model).
}
\label{fig:acomp}
\end{figure}

Figure~\ref{fig:acomp} shows the polarization degree in the $60-500$~keV
band as a function of $q$ for several values of $y_j$.
The other parameters are $\gamma\nu'_0 = 350$~keV, $\alpha=-0.2, \beta=1.2,$
and $z=1$.
The results of our calculations for the 
case of $\alpha=\beta$ are consistent with those of \citet{lazzati04}.
The results are similar to those of the SR model, but the polarization
degree is higher than in the SR model.

The polarization is higher for softer spectra,
although the local polarization degree is not dependent on the frequency
in this model.
For instance, for $y_j = 1$, $\gamma\nu'_0=350$~keV, and $z=1$,
the maximum polarization is 
$\simeq 0.66$ for $\alpha=-0.5$ and $\beta=0.9$, 
while it is $\simeq 0.71$ for $\alpha=0.4$ and $\beta=1.8$,
but the variation is smaller than for the synchrotron models 
(see \S~\ref{subsec:so} and \ref{subsec:sr}).
This variation is caused by the kinematic effect.
The local polarization degree is a maximum for $\theta = \gamma^{-1}$
(i.e., $\theta' = \pi/2$).
Thus the net polarization is higher when the contribution of the emission
from higher latitude with $\theta > \gamma^{-1}$ is smaller.
The high latitude emission is dimmer as the radiation spectrum is softer.
Therefore the net polarization is higher when the spectrum is softer.
This effect also arises in the SO and SR models, although in those
models the intrinsic dependence of polarization on the spectrum 
(equation~\ref{eq:synch_pi}) is rather strong 
(see \S~\ref{subsec:so} and \ref{subsec:sr}).
For $y_j \geq 0.01$, $\alpha \leq 0.4$, and $\beta \leq 1.8,$
the maximum polarization degree for the CD model is $\simeq 1.0$.

\section{Statistical properties}
\label{sec:simulation}

In this section we show the results of our Monte Carlo simulation of the 
GRB prompt emission polarization.
First, in \S~\ref{subsec:flu},
we give the values of the model parameters so that the observed fluences
and peak energies of simulated bursts are consistent with the data obtained
with the {\it HETE-2} satellite.
In \S~\ref{subsec:sum},
we examine the properties of the polarization distribution
of bursts detectable by the {\it POET} satellite, regardless of instrument MDP.
Next, in \S~\ref{subsec:cum},
we show the distribution of polarizations that can be measured by 
{\it POET}, and discuss how we may constrain the emission models.

\subsection{Model parameters}
\label{subsec:flu}

We performed Monte Carlo simulations to obtain the distribution
of the observed spectral energies and fluences in the three emission models.
Such simulations have been developed to discuss 
the empirical correlation
between spectral peak energies in the cosmological rest frame
and isotropic $\gamma$-ray energies
among GRBs and X-ray flashes 
in several models of geometrical structure of GRB jets
\citep{zhang04,yamazaki04,dai05,toma05,donaghy06}.
We generated 10,000 GRB jets with Lorentz factor, $\gamma$, and opening
angle, $\theta_j$, and a random viewing angle for each jet according to
the probability distribution of $\sin\theta_v d\theta_v d\phi$
with $\theta_v < 0.22$~rad.
\footnote{
We confirmed that the bursts with $\theta_v \geq 0.22$~rad in our simulation are 
not detected by
{\it HETE-2} or {\it POET} with the parameters we adopt
in this paper.
We can therefore 
discuss the distribution of several quantities of the detectable bursts
and the event rate ratio of bursts for which polarizations can be measured
to the detectable bursts without considering the bursts with $\theta_v \geq 0.22$~rad.
}
For each burst generated we calculate 
the $\nu I_{\nu}$ spectrum to obtain
the spectral peak energy,
$E_{p,{\rm obs}}$, and the fluence, $I$, in the $2-400$~keV range
by using equation (\ref{eq:flu}).
Since $E_{p,{\rm obs}}$'s and $I$'s calculated for each $q=\theta_v/\theta_j$ in
the three models are different only by factors less than 2,
$E_{p,{\rm obs}}$'s and $I$'s of the simulated bursts may be calculated using 
just one model, for which we chose the CD model. 

\begin{figure}
\epsscale{1.2}
\plotone{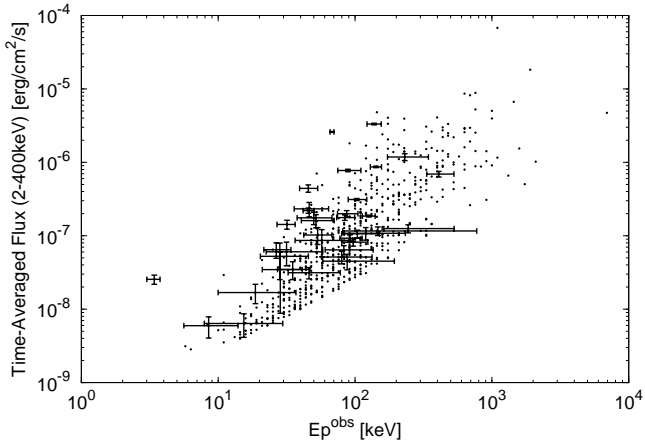}
\caption{
The $E_{p, {\rm obs}}-F$ diagram calculated in our Monte Carlo simulation.
The simulated events that can be detected by WXM on {\it HETE-2}
are represented by dots.
They are compared with the {\it HETE-2} data ({\it points with errorbars})
\citep{sakamoto05}.
}
\label{fig:ep_flu}
\end{figure}

The distributions of $\gamma$ and $\theta_j$ for GRB jets are highly uncertain.
We make a simple assumption for the distribution and in \S~\ref{subsec:cum}
we perform some simulations for different assumptions.
We fix $\gamma=100$.
We assume the distribution of $\theta_j$ as
\begin{equation}
f(\theta_j) d\theta_j \propto \left\{
\begin{array}{lr}
\theta_j^{q_1} d\theta_j &{\rm for}~0.001 \leq \theta_j \leq 0.02 \\
\theta_j^{q_2} d\theta_j &{\rm for}~0.02 \leq \theta_j \leq 0.2,
\end{array}
\right.
\label{eq:thetaj}
\end{equation}
where $q_1 = 0.5$ and $q_2 = -2.0$.
The value of $q_2=-2$ is inferred from the 
observations of the steepening breaks (i.e., jet breaks)
of some optical afterglows \citep{frail01,zeh06}
and from analysis of {\it BATSE} data using some empirical relations
\citep{yonetoku05}.
There are several suggestions of events with very small $\theta_j$
\citep[e.g.,][]{schady07,racusin08},
although the value of $q_1$ is highly uncertain.
The spectral parameters $r_0^2 A_0, \gamma \nu'_0, \alpha,$ and $\beta$ are
assumed as follows.
The first two parameters are given so that the rest-frame spectral peak
energies and isotropic $\gamma$-ray energies calculated for a jet
viewed with $\theta_v = 0$ are consistent with those of typical GRBs.
Such an on-axis emission has approximately 
$E_p = 2\gamma\nu'_0$ and $E_{\rm iso} = 16\pi^2 r_0^2 A_0 \gamma \nu'_0$.
The parameters $r_0^2 A_0$ and $\gamma\nu'_0$ are given through the
empirical relations $E_{\rm iso} \theta_j^2/2 = 10^{51} \xi_1$~erg
and $E_p = 80 \xi_2 (E_{\rm iso}/10^{52}~{\rm erg})^{1/2}$~keV
\citep[e.g.,][]{frail01,amati02}.
We assume that the coefficients $\xi_1$ and $\xi_2$ obey the log-normal
distribution \citep{ioka02} with averages of $1$ and logarithmic variances
of $0.3$ and $0.15$, respectively.
The last two parameters are fixed by $\alpha=-0.2$ and $\beta=1.2$, which
are typical values for GRB prompt emission \citep{preece00,sakamoto05}.
The distribution of the source redshift, $z$, 
is assumed to be in proportional to the cosmic star formation rate.
We adopt the model SF2 in \citet{porciani01}, 
i.e., the comoving GRB rate density is assumed to be proportional to
\begin{equation}
R(z) = \frac{\exp(3.4z)}{\exp(3.4z)+22} 
\frac{\sqrt{\Omega_M (1+z)^3 + \Omega_{\Lambda}}}{(1+z)^{3/2}}.
\end{equation}
We take the standard cosmological parameters of 
$H_0 = 70~{\rm km}~{\rm s}^{-1}~{\rm Mpc}^{-1}, 
\Omega_M = 0.3,$ and $\Omega_{\Lambda} = 0.7$.

Figure~\ref{fig:ep_flu} shows the results of $E_{p, {\rm obs}}$ and 
time-averaged flux, $F$.
The time-averaged flux is calculated by $F=I/T$, where $T$ is the duration
of a burst.
We fix $T=20$~s, which is a typical value for long GRBs \citep[e.g.,][]{sakamoto05}.
We show only the simulated bursts that have fluxes above the detectable
limit of the {\it HETE-2} satellite.
They are consistent with the data obtained by {\it HETE-2} 
\citep{sakamoto05}.
The scatter of the simulated bursts is due both to the scatter of the 
assumed jet parameters and the viewing angle effect
\citep{yamazaki04,donaghy06}.
\footnote{
\citet{yamazaki04} showed a deviation from the Amati relation 
($E_p \propto E_{\rm iso}^{1/2}$) for $E_p < 10$~keV in the uniform jet model,
but the $E_{p, {\rm obs}}-F$ diagram we derive is still consistent with 
the observed dataset.
}

\subsection{Properties of polarization distribution}
\label{subsec:sum}

\begin{figure}
\epsscale{1.2}
\plotone{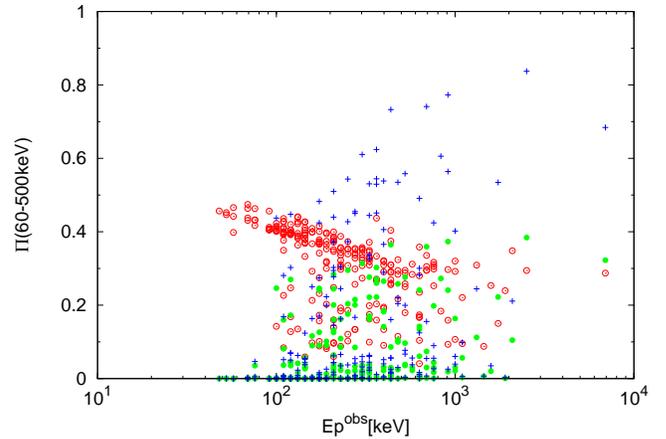}
\caption{
$E_{p, {\rm obs}}-\Pi$ diagrams for the simulated events that can be
detected by GRAPE in the
SO ({\it red open circles}), SR ({\it green filled circles}), and 
CD ({\it blue plus signs}) models.
The adopted parameters are as follows.
The fixed parameters are $\gamma=100, q_1=0.5, q_2=-2.0, \alpha=-0.2,$
$\beta=1.2,$ and $T=20$~s.
The distribution of the source redshift $z$ is assumed to be in proportional
to the cosmic star formation rate.
The parameters $r_0^2 A_0$ and $\gamma \nu'_0$ are distributed so that
the simulated $E_{p,{\rm obs}} - F$ diagram is consistent with the observed 
data (see Figure~\ref{fig:ep_flu}).
See text for the cases of the spectral indices distributed realistically, for
$-0.5<\alpha<0.4$ and $0.9<\beta<1.8$.
}
\label{fig:ep_pi_1_g}
\end{figure}

\begin{figure}
\epsscale{1.2}
\plotone{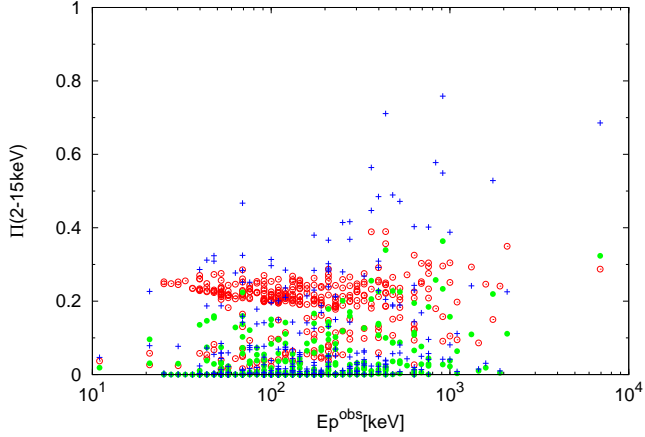}
\caption{
Same as Figure~\ref{fig:ep_pi_1_l}, but for LEP.
}
\label{fig:ep_pi_1_l}
\end{figure}

We calculated the linear polarization, $\Pi$, by using equations
(\ref{eq:atorob}), (\ref{eq:arndb}), and (\ref{eq:acomp})
to obtain the polarization distribution of the simulated bursts that
can be detected by GRAPE and LEP.
The detection limits of GRAPE and LEP are set to be the MDP contours of 1.0.
(see Figure~\ref{fig:mdp})
\footnote{
The detection limits of GRAPE and LEP 
for signal-to-noise ratio $>5$ are similar but not
identical to the MDP contours of 1.0. 
Thus our setting for the detection limits is just for simplicity.
}. 
Figures~\ref{fig:ep_pi_1_g} and \ref{fig:ep_pi_1_l}
show the $E_{p,{\rm obs}}-\Pi$ diagrams of all the simulated bursts that
can be detected by GRAPE and LEP, respectively,
in the SO ({\it red open circles}), 
SR ({\it green filled circles}), and CD ({\it blue plus signs}) models.
In the SO model, most of the detectable bursts have $0.3<\Pi<0.5$ in
the GRAPE band (60-500~keV), while they have $0.2<\Pi<0.3$ in the LEP
band (2-15~keV).
In the SR and CD models, most of the detectable bursts have $\Pi<0.1$
in both GRAPE and LEP bands.
The events with $\Pi \geq 0.1$ are distributed uniformly with $\Pi < 0.4$ and
$\Pi < 0.9$ for the SR and CD models, respectively.

\begin{figure}
\epsscale{1.2}
\plotone{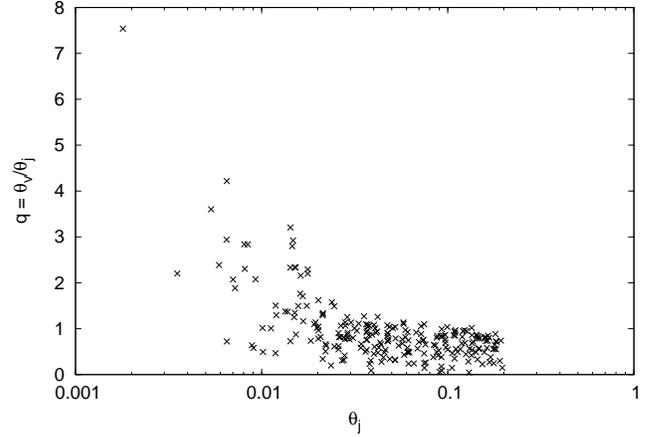}
\caption{
Distribution of $\theta_j$ and $q=\theta_v/\theta_j$ of the detectable
bursts by GRAPE in the model described by Figure~\ref{fig:ep_pi_1_g}.
}
\label{fig:thj_q}
\end{figure}

This result can be roughly explained by the polarizations calculated as functions of
$y_j$ and $q=\theta_v/\theta_j$ for $\gamma \nu'_0 = 350$~keV and $z=1$
(see Figures~\ref{fig:atorob}, \ref{fig:arndb}, \ref{fig:acomp}) and
the distribution of $\theta_j$ and
$q$ for the detectable bursts in this simulation, shown in Figure~\ref{fig:thj_q}. 
The detectable events are dominated by the events with $q<1$,
since events with $q<1$ are much brighter than those with $q>1$ because
of the relativistic beaming effect.
For events with $q>1$, narrower jets are easier to detect since
they have intrinsically higher emissivities by our assumption.
Most of the detectable events have $q<1$ and $\theta_j > 0.02$ (i.e., $y_j > 4$).
For these events the SO model gives $0.3<\Pi<0.5$ in most cases,
while the SR and CD models give $\Pi<0.1$, for the GRAPE band as shown in Figures~
\ref{fig:atorob}, \ref{fig:arndb}, and \ref{fig:acomp}.
The remaining detectable events mainly have $q>1$ and $\theta_j>0.005$ 
(i.e., $y_j>0.25$).
These events have $\Pi < 0.6$ in the SO model, 
$\Pi < 0.5$ in the SR model, and $\Pi < 0.9$ in the CD model,
for the GRAPE band as shown in Figures~
\ref{fig:atorob}, \ref{fig:arndb}, and \ref{fig:acomp}.
The results for the LEP band can be explained similarly.

In all the three models,
the results show $\Pi(60-500~{\rm keV})>\Pi(2-15~{\rm keV})$ for 
almost all the detectable bursts with $\Pi>0.1$.
This is due to the fact that 
typically the contribution of the high-energy photons with spectral index $\beta$
is larger in the GRAPE band than in the LEP band.
The emission with softer spectrum has higher polarization because of
the intrinsic property of the synchrotron polarization 
(equation \ref{eq:synch_pi}) for the SO and SR models and the kinematic
effect for the CD model (see \S~\ref{subsec:compton}), respectively.

In the SO model, the polarization of GRBs with $q<1$ is higher
for lower $E_{p,{\rm obs}}$ for the GRAPE band.
This is because the contribution from high-energy
photons, with energy spectral index $\beta$, is larger.
In the SR and CD models,
the higher $\Pi$ GRBs can be obtained for smaller $\theta_j$.
The maximum $\Pi$ is obtained for $\theta_j \simeq 0.002$.

\subsection{Cumulative distribution of measurable polarizations}
\label{subsec:cum}

\begin{figure}
\plottwo{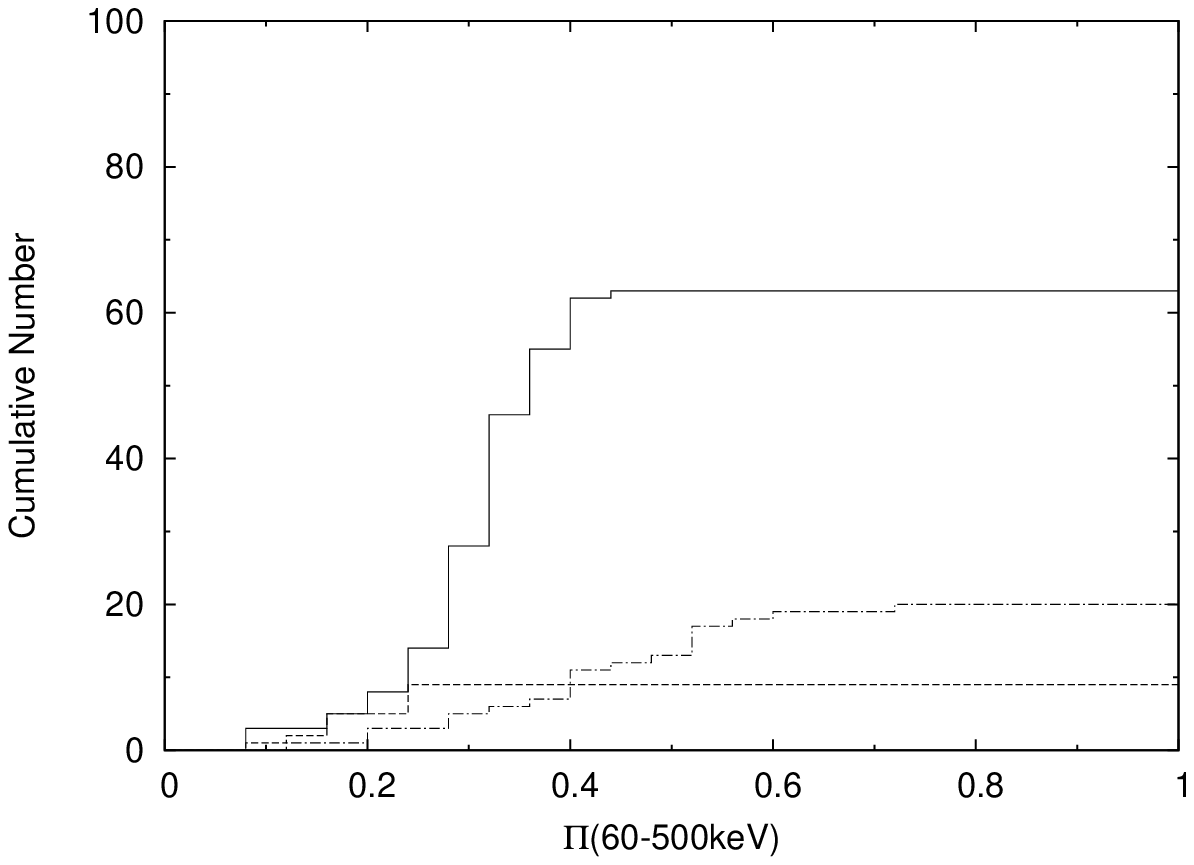}{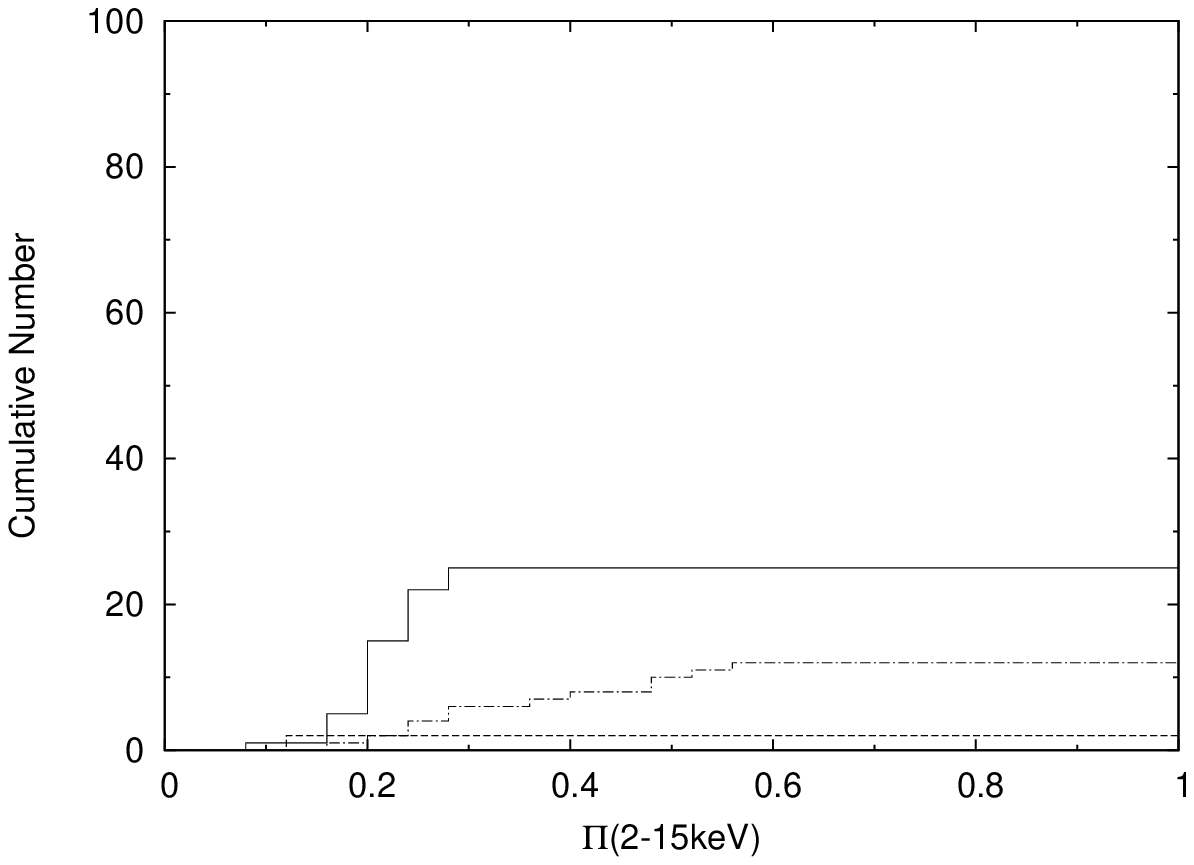}
\caption{
The cumulative distribution of $\Pi$ that can be measured by GRAPE
({\it left}) and LEP ({\it right}) in the SO ({\it solid}), 
SR ({\it dashed}), and CD ({\it dot-dashed}) models in which
the number of detectable bursts is 200.
The adopted parameters are as follows.
The fixed parameters are $\gamma=100, q_1=0.5, q_2=-2.0, \alpha=-0.2,$
$\beta=1.2,$ and $T=20$~s.
The distribution of the source redshift $z$ is assumed to be in proportional
to the cosmic star formation rate.
The parameters $r_0^2 A_0$ and $\gamma \nu'_0$ are distributed so that
the simulated $E_{p,{\rm obs}} - F$ diagram is consistent with the observed 
date (see Figure~\ref{fig:ep_flu}).
See text for the cases of
the spectral indices distributed realistically, for $-0.5<\alpha<0.4$ and
$0.9<\beta<1.8$.
}
\label{fig:cum_1}
\end{figure}

We obtain the distribution of polarization that can be measured, by using
the MDP values we derived for 
$\alpha=-0.2$, $\beta=1.2$, and $T=20$~s (see \S~\ref{sec:poet}).
We interpret the simulated events with $\Pi > MDP$ as 
`$\Pi$-measurable events'.
Figure~\ref{fig:cum_1} shows the cumulative distribution of $\Pi$ that
can be measured by GRAPE and LEP in the SO, SR, and CD models.
We have set the number of detectable events $N_d=200$.
Since the polarization in the LEP band is lower than in the
GRAPE band for almost all the cases as discussed in \S~\ref{subsec:sum},
the number of events for which
polarization can be measured by LEP is smaller than for GRAPE.
In the SO model, the number of $\Pi$-measurable bursts is 
$N_m > 60$, and  
the cumulative distribution of measurable $\Pi$
varies rapidly at $0.3<\Pi<0.4$ for the GRAPE band.
In the SR model, $N_m < 10$, and the maximum polarization is
$\Pi_{\rm max} < 0.4$.
In the CD model, $N_m < 30$, and $\Pi_{\rm max} < 0.8$.

To investigate general properties of the cumulative distribution that 
do not depend on the model parameters,  
we performed simulations for other values of $\gamma,$ $q_1,$ and $q_2$,
the Lorentz factor of the jets and the power-law indices of the distribution of the 
opening angles of the jets, respectively.
We refer to the parameters adopted for the above simulation as `typical' parameters.
We now consider a range of parameters: 
$\gamma \geq 100$, $q_1 \geq 0.5,$ and $q_2 \geq -3.0$, which
are quite reasonable for GRBs \citep[e.g.,][]{lithwick01,yonetoku05}.
Within these parameter ranges we obtain the lower (upper) limit
of $N_m/N_d$ for the SO model (the SR/CD models).

\begin{figure}
\plottwo{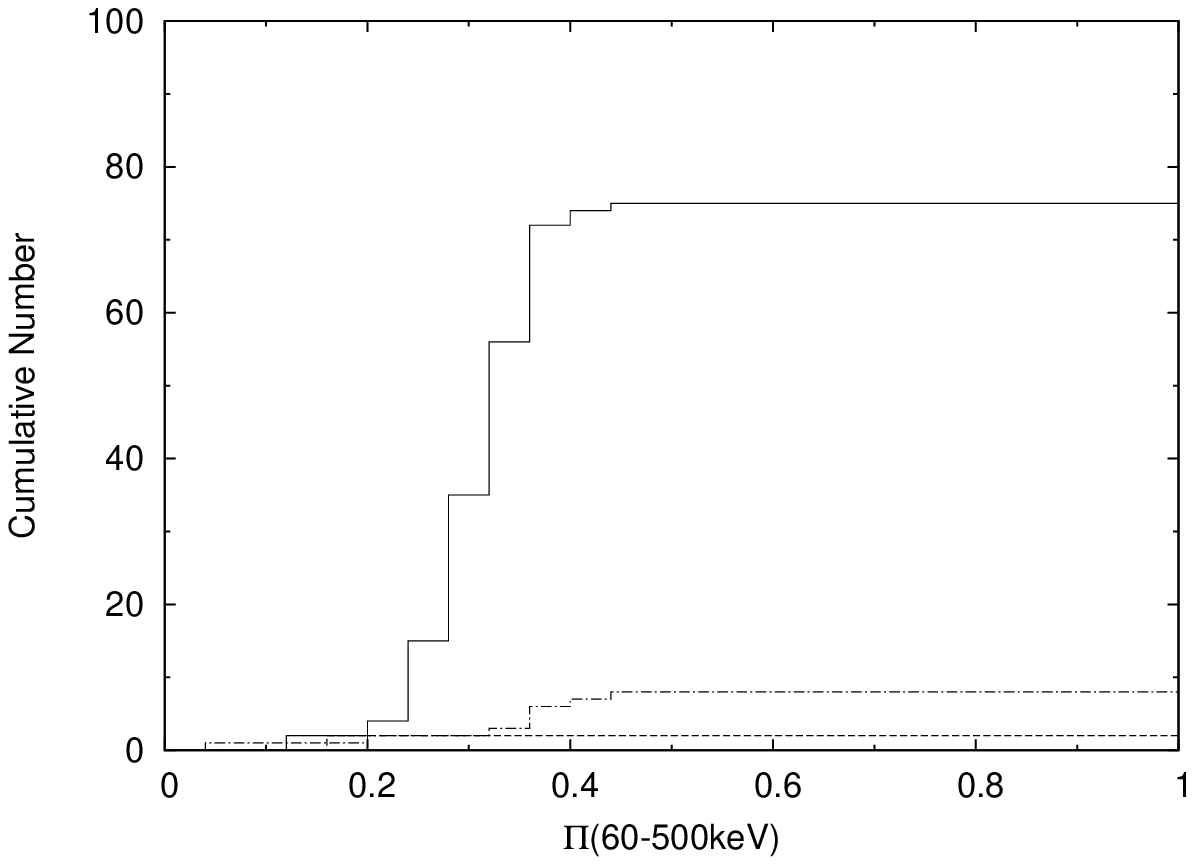}{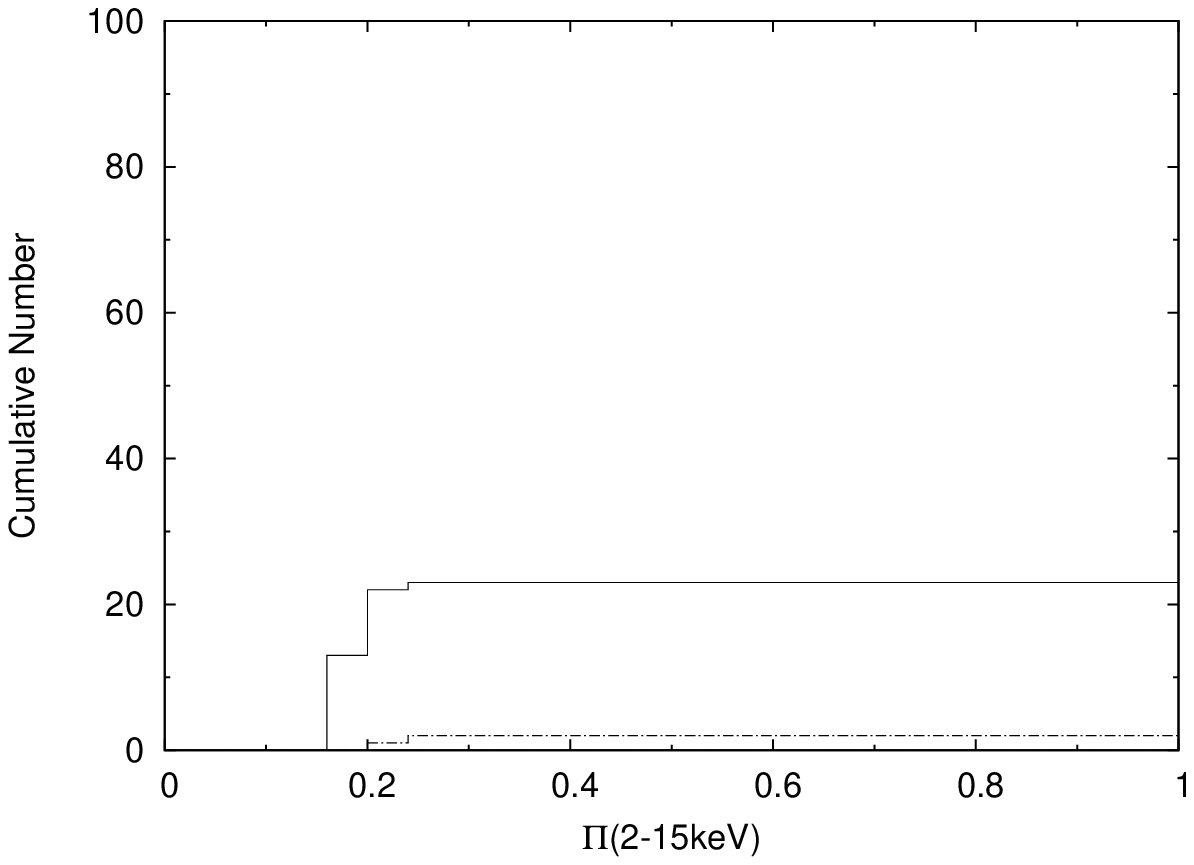}
\caption{
Same as Figure~\ref{fig:cum_1}, but the Lorentz factor
of the jets $\gamma=300$.
}
\label{fig:cum_g}
\end{figure}

\begin{figure}
\plottwo{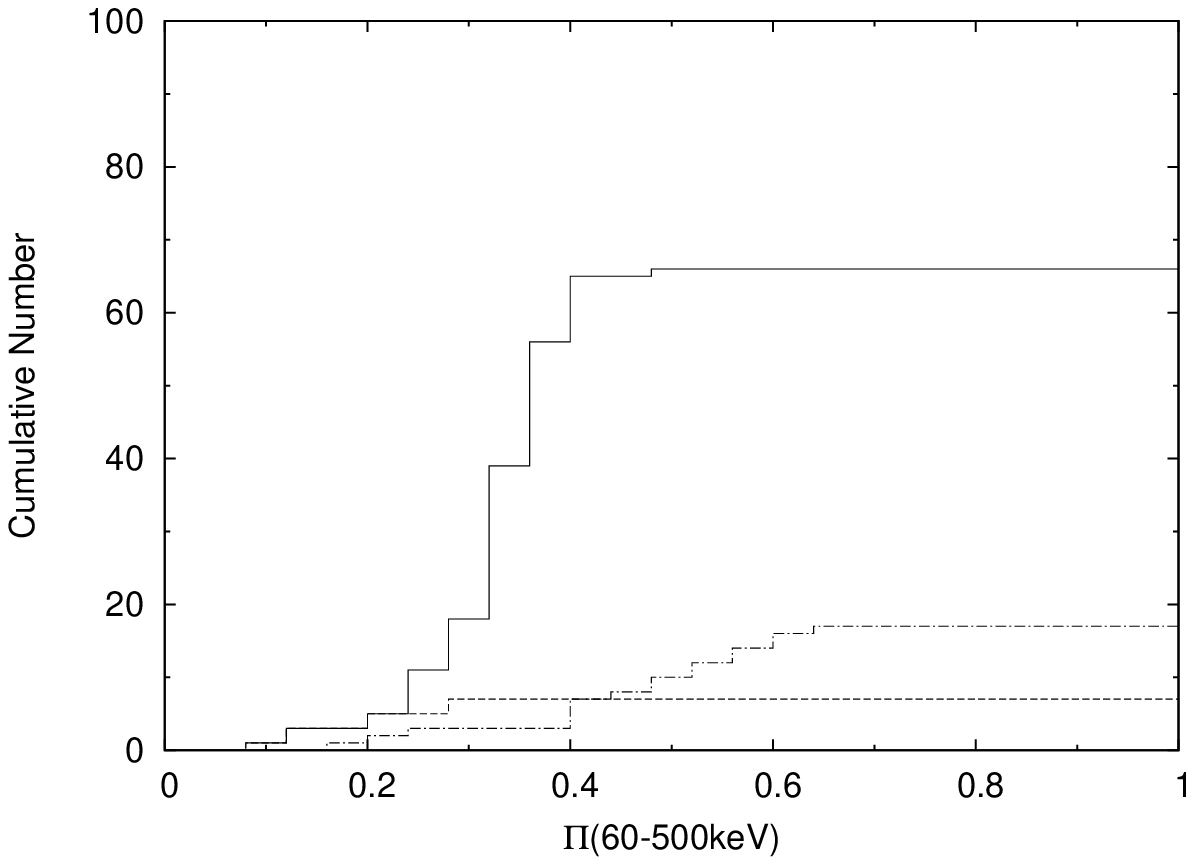}{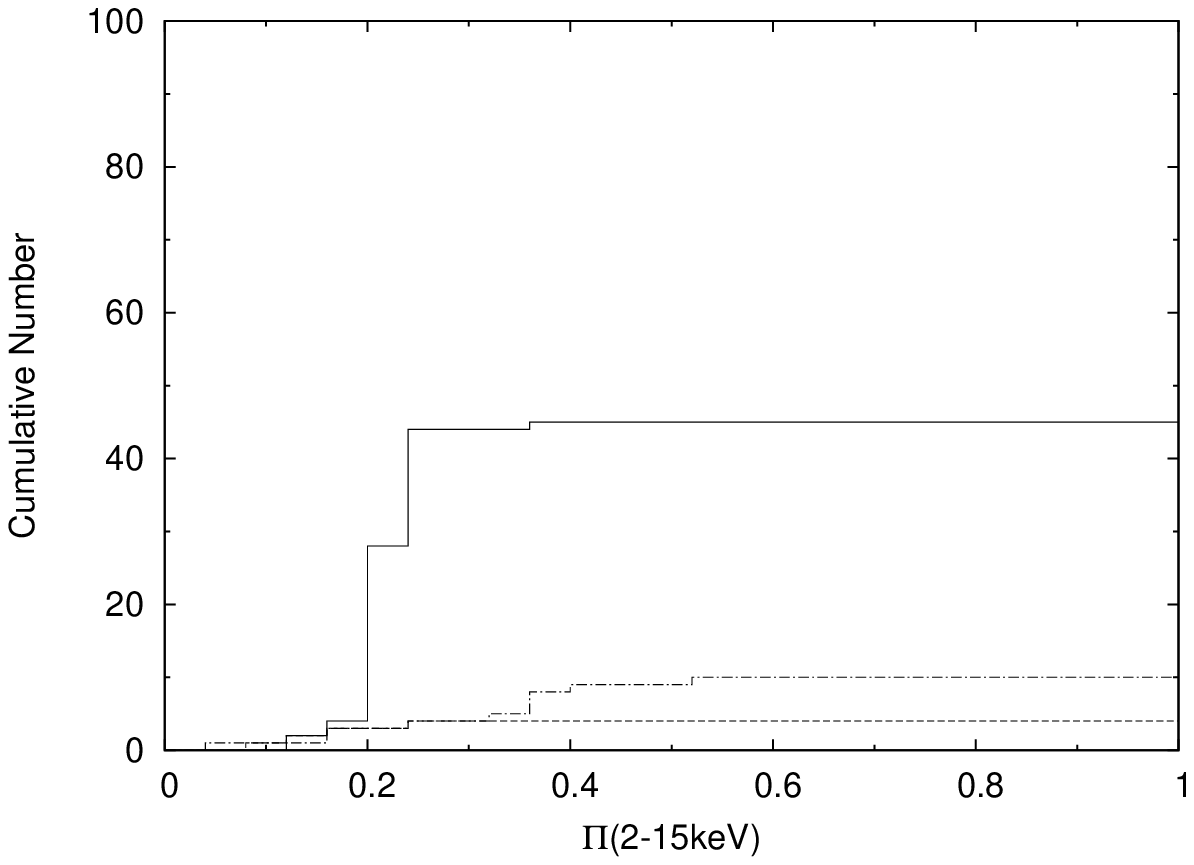}
\caption{
Same as Figure~\ref{fig:cum_1}, but the lower power-law
index of the $\theta_j$ distribution
$q_1 = 1.0$ (see equation \ref{eq:thetaj}).
}
\label{fig:cum_q1}
\end{figure}

\begin{figure}
\plottwo{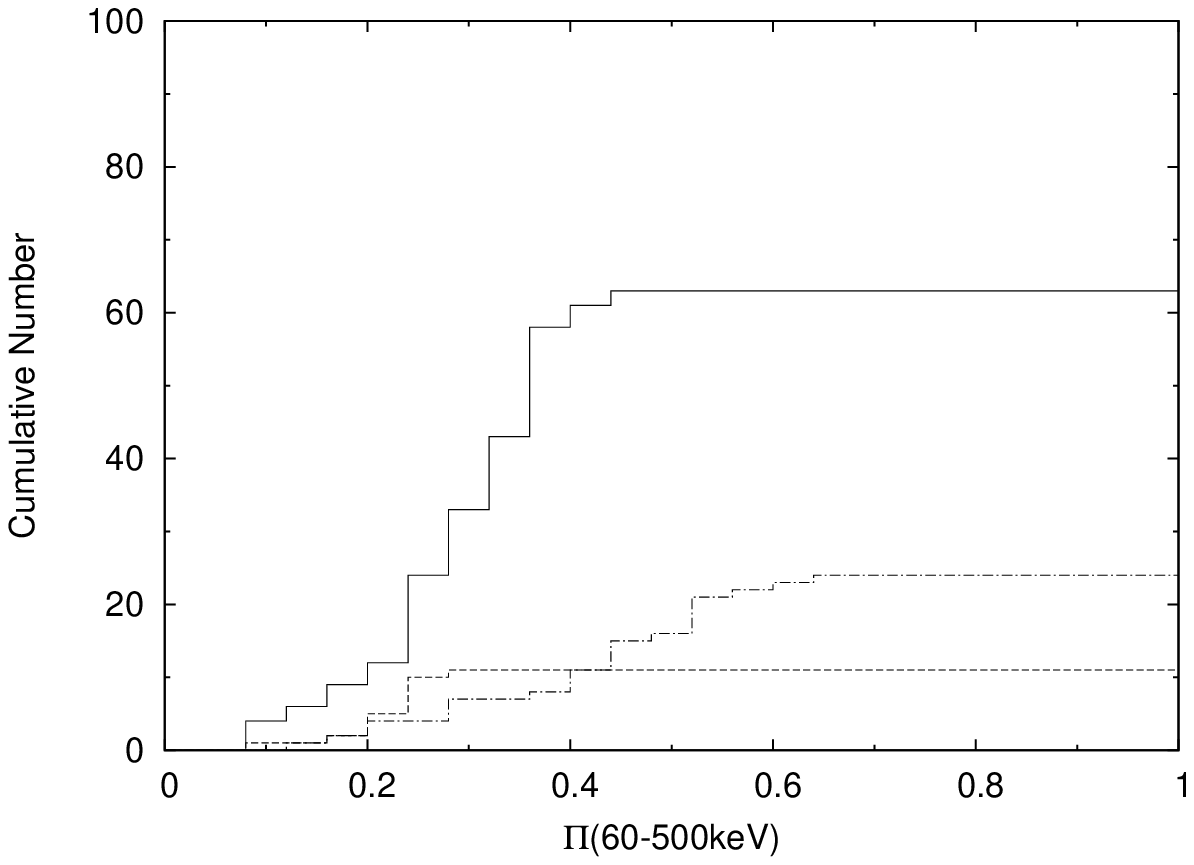}{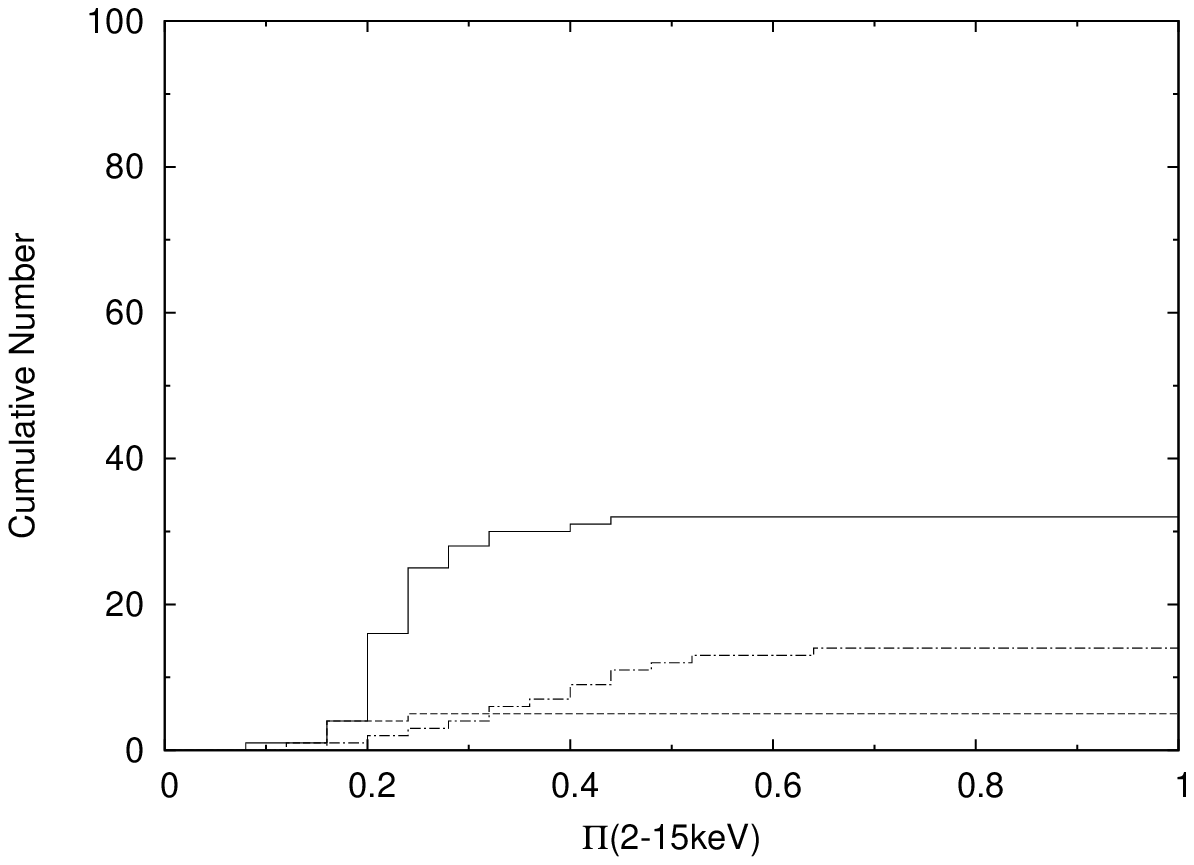}
\caption{
Same as Figure~\ref{fig:cum_1}, but the higher power-law
index of the $\theta_j$ distribution
$q_2 = -3.0$ (see equation \ref{eq:thetaj}).
}
\label{fig:cum_q2}
\end{figure}

Figure~\ref{fig:cum_g} shows the results for $\gamma=300$ and 
the same `typical' values for the other parameters.
The number $N_m$ is larger in the SO model and
smaller in the SR and CD models than the case of $\gamma=100$.
As $\gamma$ is larger, the beaming effect is stronger and 
the ratio of the bursts with $q<1$ for detectable bursts is larger.
Thus the number of bursts with a high degree of polarization is larger
in the SO model and smaller in the SR and CD models.
Figure~\ref{fig:cum_q1} shows the results for $q_1=1.0$ and the same
`typical' values for the other parameters.
Since the ratio of the number of the bursts with smaller $y_j$ to
that of detectable bursts is smaller, 
$N_m$ is slightly smaller than that for the `typical' parameters 
in the SR and CD models.
Figure~\ref{fig:cum_q2} shows the results for $q_2=-3.0$ and the `typical'
values for the other parameters.
In this case $N_m$ is slightly larger
than that for the `typical' parameters in the SR and CD models.
The number $N_m$ in the SO model is similar for Figure~\ref{fig:cum_1},
\ref{fig:cum_q1}, and Figure~\ref{fig:cum_q2} in the GRAPE band.
To summarize, for the parameters $\gamma\geq 100$, $q_1 \geq 0.5$, $q_2 \geq -3.0$,
$\alpha=-0.2$ and $\beta=1.2$, we can say that
$N_m/N_d > 30\%$ for GRAPE
and the cumulative distribution of
measurable $\Pi$ varies rapidly from $0.3<\Pi<0.4$ in the SO model.
For the SR model, $N_m/N_d < 5\%$ for GRAPE, with a maximum polarization
$\Pi_{\rm max} < 0.4$.
For the CD model, $N_m/N_d < 15\%$ for GRAPE,
and $\Pi_{\rm max} < 0.8$.

Since the dependence of the polarization degree on the spectral indices
is relatively large in the SO and SR models, we should take account of 
the distribution of $\alpha$ and $\beta$.
The observed spectral parameters $\alpha$ and $\beta$ are distributed
roughly as $-0.5<\alpha<0.4$ and $0.9<\beta<1.8$ \citep{preece00,sakamoto05}.
Within these ranges of $\alpha$ and $\beta$,
the polarization degree for
$y_j>10$, $q<1$, and $50<E_{p,{\rm obs}}<10^3$~keV is $0.2<\Pi<0.7$ in the SO model.
Thus the measurable polarizations are clustered at $0.2<\Pi<0.7$.
The maximum polarization obtained in the SO model 
for $y_j \geq 0.01, \alpha \leq 0.4,$
and $\beta \leq 1.8$ is $\simeq 0.8$ (see \S~\ref{subsec:so}).
In this case $N_m/N_d$ will be larger than 30\%.
In the CD model, the result will not be significantly different from the case of 
fixed $\alpha$ and $\beta$.
In the SR model, the polarization degree does not exceed those calculated in
the CD model, and thus $N_m/N_d < 15\%$.
The maximum polarization obtained in the SR model 
for $y_j \geq 0.01, \alpha \leq 0.4,$
and $\beta \leq 1.8$ is $\simeq 0.8$ (see \S~\ref{subsec:sr}).

In conclusion, we can constrain the emission mechanism of GRBs by using the
cumulative distribution obtained by GRAPE.
If $N_m/N_d > 30\%$, the SR and CD models may be ruled out, and in this case
if the measured polarizations are clustered at $0.2<\Pi<0.7$,
the SO model will be favored.
If $N_m/N_d < 15\%$,
the SO model may be ruled out,
but we cannot distinguish between the SR and CD models with different 
distributions of $y_j$, $\alpha$, and $\beta$.
If several bursts with $\Pi > 0.8$ are detected, however, the CD model
which includes adequate number of small $y_j$ bursts will be favored.

\section{Summary and discussion}
\label{sec:summary}

Recently there has been an increasing interest in the measurement of 
X-ray and $\gamma$-ray polarization, and
the observational techniques can now achieve significant sensitivity 
in the relevant energy bands.
Several polarimetry mission concepts, such as {\it POET}, are being planned.
The {\it POET} concept has two polarimeters, GRAPE (60-500~keV)
and LEP (2-15~keV) both of which have wide fields of view.
If launched, missions of this type would
provide the first definitive detection of the
polarization of GRB prompt emission. 
This would enable the discussion of the statistical properties of the 
polarization degree and polarization spectra, which
will give us diagnostic information on the emission mechanism
of GRBs and the nature of the GRB jets that cannot be obtained 
from current spectra and lightcurve observations.
We have performed Monte Carlo simulations of the linear polarization
from GRB jets for three major emission models: 
synchrotron model with globally ordered magnetic field (SO model), 
synchrotron model with small-scale random magnetic field (SR model), 
and Compton drag model (CD model).
We assumed that the physical quantities for the emission of the jets
are uniform on the emitting surface and that the jets have sharp edges.
Our jet angle distribution allows the detections of
GRBs with very small opening angles (i.e., smaller than
1 degree) as suggested by several {\it Swift} bursts \citep{schady07,racusin08}.
We have shown that the {\it POET} mission 
or other polarimeters with similar capabilities,
i.e., broadband spectral capabilities for the determination of $E_{p,{\rm obs}}$ 
and sensitive broadband polarimetric capabilities to minimize MDP,
can constrain the emission models of GRBs. 
Furthermore, these simulations indicate that an increase in the LEP effective area
would be beneficial to compensate for the lower expected polarization 
at lower energies.

As shown in Figures~\ref{fig:atorob}, \ref{fig:arndb}, and \ref{fig:acomp},
the SR and CD models require off-axis observations of the jets
to achieve a high level of polarization, while the SO model does not.
In this sense the SR and CD models are categorized as {\it geometric}
models, and the SO model as an {\it intrinsic} model
\citep{waxman03,lazzati06}.
The distribution of observed polarizations obtained by our simulations
show that the geometric SR/CD models will be ruled out
if the number ratio of the $\Pi$-measurable bursts to detected bursts
is larger than $30\%$,
and in this case the SO model will be favored
if the measurable polarizations are clustered at $0.2<\Pi<0.7$.
If the number ratio is smaller than $15\%$,
the SO model may be ruled out,
but we cannot distinguish between the SR and CD models with different 
distributions of $y_j=(\gamma \theta_j)^2$, $\alpha$, and $\beta$,
where $\gamma$ and $\theta_j$ are the bulk Lorentz factor and the
opening angle of the GRB jet, respectively, and $\alpha$ and $\beta$
are lower and higher indices of the energy spectrum.
However, if several bursts with $\Pi > 0.8$ are detected, the CD model
which includes an adequate number of small $y_j$ bursts will be favored.

If the cumulative distribution of the measurable polarizations favors 
the SO model,
the globally ordered magnetic field would be advected from the central engine.
If we understand the strength of the magnetic field in the emitting 
region from the luminosity and the spectrum of the emission,
we can constrain the strength of the field at the central engine.
If the geometric SR/CD models are favored from the observations, 
it will be established, independently of the afterglow observations, 
that GRB outflows are not spherical but highly collimated.
If the CD model is favored by the observations, we may constrain the 
distribution of the parameter $y_j = (\gamma \theta_j)^2$ of GRB jets.
The CD model needs a dense optical/UV photon field 
interacting within the relativistic jets \citep{lazzati00,eichler03}.

We have made some simplifications in our simulations, and there are some caveats.
We have assumed that the jets are uniform on the emitting surfaces and 
have sharp edges.
To compare the simulations and the observations further, more sophisticated
modeling is required \citep[e.g.,][]{zhang04,toma05}.

We have interpreted bursts as a simple combination of pulses, without taking
account of the temporal variation of the Lorentz factor $\gamma$ of the jet.
If this is accounted for,
each pulse may have different $y_j=(\gamma \theta_j)^2$ but the same 
$q=\theta_v/\theta_j$.
We should then average the polarization with respect to fluence of each
pulse having different $y_j$ \citep{granot03,nakar03}.
However, in the SO model, the cumulative distribution of measurable $\Pi$
will not be changed significantly as long as $y_j > 10$, because
$\Pi$ is clustered into a small range for $q<1$ and $y_j>10$.
To average the polarization in the case of $q>1$, the relation between
the luminosity and the Lorentz factor for each pulse is required
to predict the polarization distribution.

For the SR model we have assumed that the directions of the magnetic field
are confined within the shock plane.
They may be more isotropic in reality, in which case the polarization degree
in the SR model will be reduced.

In the synchrotron model with a combination of the globally ordered magnetic
field and the locally random field, 
$\mathbf{B} = \mathbf{B}_{\rm ord} + \mathbf{B}_{\rm rnd}$,
the linear polarization can be calculated by 
$\Pi = (Q_{\rm ord} + Q_{\rm rnd})/(I_{\rm ord} + I_{\rm rnd})
\approx (\Pi_{\rm ord}+\eta \Pi_{\rm rnd})/(1+\eta)$,
where $\{I,Q\}_{\rm ord}$ and $\{I,Q\}_{\rm rnd}$ are the Stokes parameters
from the ordered and random fields, respectively. 
$\Pi_{\rm ord}$ and $\Pi_{\rm rnd}$ are described by equations 
(\ref{eq:atorob}) and (\ref{eq:arndb}), and 
$\eta \equiv (B_{\rm rnd}/B_{\rm ord})^{\alpha+1}$.
This model will reduce
the number ratio of $\Pi$-measurable bursts to detected bursts to
less than 30\% and 
the clustering of measurable polarizations will be at $\Pi < 0.7$.

\acknowledgements
This work is supported in part by
the Grant-in-Aid from the 
Ministry of Education, Culture, Sports, Science and Technology
(MEXT) of Japan, No.19047004 (RY, KI, and TN), 
No.18740153 (RY), No.18740147 (KI),
and in part by the Grant-in-Aid for the global COE program
{\it The Next Generation of Physics, Spun from Universality and Emergence}
from the Ministry of Education, Culture, Sports, Science and
Technology (MEXT) of Japan.
BZ acknowledges NASA NNG05GB67G and NNX08AE57A (Nevada NASA EPSCoR 
program) and KT acknowledges NASA NNX08AL40G for partial supports.

\appendix

\section{Some notes on synchrotron polarization}

\subsection{The SO model: synchrotron with ordered field}
\label{subsec:app_so}

We consider the synchrotron radiation from the shell moving radially outward
with a bulk Lorentz factor $\gamma \gg 1$, and 
the magnetic field in the shell is globally ordered within the plane
parallel to the shock plane.
If the matter of the shell expands with a constant speed, the strength
of magnetic field with radial direction scales as $R^{-2}$ while that with
transverse direction scales as $R^{-1}$.
Thus the field advected with the shell is likely to have the direction
parallel to the shock plane.

We set the line of sight
(i.e., the direction from the central engine to the earth) in the lab frame
to be $z$ axis,
and the direction of the magnetic field on a given point
of the shell, projected onto the plane perpendicular to $z$ axis,
to be $\bar{x}$ axis.
The given point can be described by spherical coordinates $(\theta, \varphi)$.
Then the components of the velocity vector of the given point and 
the unit wave vector can be described by the right-handed coordinate system
$\bar{x} \bar{y} z$ as
$\mathbf{\beta} = (\beta \sin\theta \cos\varphi, \beta\sin\theta \sin\varphi,
\beta\cos\theta)$ and $\hat{\mathbf{k}} = (0, 0, 1)$, respectively.
The unit wave vector in the comoving frame is
\begin{equation}
\hat{\mathbf{k}'} = \frac{1}{\gamma (1-\mathbf{\beta} \cdot \hat{\mathbf{k}})}
\left[ \hat{\mathbf{k}} + \mathbf{\beta} \left( \frac{\gamma^2}{\gamma+1}
\mathbf{\beta}\cdot\hat{\mathbf{k}} - \gamma \right)\right].
\end{equation}
Since the direction of the magnetic field in the comoving frame is 
perpendicular to the velocity vector of the fluid,
$\hat{\mathbf{B}}' = (\cos\theta / \sqrt{\cos^2\theta + 
\sin^2\theta \cos^2\varphi}, 0, -\sin\theta \cos\varphi/
\sqrt{\cos^2\theta + \sin^2\theta \cos^2\varphi})$.
Then we may calculate $\cos\theta'_B = \hat{\mathbf{B}'} \cdot 
\hat{\mathbf{k}'}$, and we obtain
\begin{equation}
\sin\theta'_B \approx \left( \frac{1-\gamma^2\theta^2}{1+\gamma^2\theta^2}
\right)^2 \cos^2 \varphi + \sin^2\varphi
\label{eq:sinthetab}
\end{equation}
in the limit $\gamma \gg 1$.

The direction of the polarization vector of the synchrotron radiation is
calculated by $\mathbf{e}' \parallel \hat{\mathbf{B}'}\times \hat{\mathbf{k}'}$.
Then we obtain the direction of the polarization vector in the lab frame by
\begin{equation}
\mathbf{e} = \gamma(1+\mathbf{\beta}\cdot\hat{\mathbf{k}'})\mathbf{e}'
- (\mathbf{\beta}\cdot\mathbf{e}')\left( \frac{\gamma^2}{\gamma+1}\mathbf{\beta}
+ \gamma \hat{\mathbf{k}'}\right).
\end{equation}
The results are $e_z = 0$ and
\begin{equation}
\tan\chi_B \equiv \frac{e_y}{e_x} = \tan\varphi - 
\frac{\beta-\cos\theta}{\beta\sin^2\theta} \frac{1}{\sin\varphi \cos\varphi}.
\end{equation}
The angle $\chi_B$ is the polarization position angle measured from the axis 
$\bar{x}$ (i.e., the direction of the local magnetic field).
The above equation can be rewritten as 
$\chi_B \approx \varphi + \arctan
[(1-\gamma^2\theta^2) \cot\varphi /(1+\gamma^2\theta^2)]$.
This result is consistent with that of \citet{granot03}.

Based on the above results, we consider the case that 
the magnetic field is axisymmetric
around the jet and has a toroidal configuration.
We set the direction from the line of sight to the jet axis to be $x$ axis.
Below we will rewrite the above results by using the azimuthal angle $\phi$
measured from $x$ axis.
In the coordinate system of $xyz$, the jet axis and the coordinates
of a given point on the shell are described as
$\mathbf{J} = (\sin\theta_v, 0, \cos\theta_v)$, and
$\mathbf{R} = (\sin\theta\cos\phi, \sin\theta\sin\phi, \cos\theta)$, respectively.
The magnetic field at the given point is given by
$\hat{\mathbf{B}'} = \mathbf{R} \times \mathbf{J}/|\mathbf{R} \times \mathbf{J}|$.
Let the unit vectors of the directions of $\mathbf{R}$ and $\hat{\mathbf{B}'}$
projected onto $xy$ plane be 
$\hat{\mathbf{r}}$ and $\hat{\mathbf{b}}$, and
$\cos\varphi = \hat{\mathbf{b}}\cdot\hat{\mathbf{r}}$.
Then we obtain
\begin{equation}
\cos^2\varphi \approx \frac{\sin^2\phi}{1+a^2-2a\cos\phi},
\label{eq:cos2phi}
\end{equation}
where $a \equiv \theta/\theta_v$.
Equation (\ref{eq:tsinthetab}) is given by inserting equation 
(\ref{eq:cos2phi}) into equation (\ref{eq:sinthetab}).
If we measure the position angle from the $x$ axis, we obtain equation 
(\ref{eq:tchi}), i.e., $\chi = \chi_B - \varphi + \phi$.
These results are consistent with those of \citet{granot05}.

\subsection{The SR model: synchrotron with random field}
\label{subsec:app_sr}

Here we consider that the directions of the magnetic fields are confined
within the plane parallel to the shock plane and that they are completely random.
This field configuration is possible if the field is generated by the shock.
In the comoving frame of the shell, we set the direction of $\hat{\mathbf{k}'}$
to be axis $3$, and set a right-handed coordinate system $123$.
Let the polar and azimuthal angles of $\hat{\mathbf{B}'}$ be
$\theta'_B$ and $\phi'_B$, respectively.
In this coordinate system, the Stokes parameters of synchrotron emissivity
are given by
\begin{equation}
j'^Q_{\nu'} = - j'^I_{\nu'} \Pi_0 \cos(2\phi'_B), ~~
j'^U_{\nu'} = - j'^I_{\nu'} \Pi_0 \sin(2\phi'_B).
\end{equation}
Next we set another right-handed coordinate system $xyz$ of which $z$ axis
is along the velocity vector of the fluid and $xz$ plane includes $\mathbf{k}'$.
Then the angle between $\mathbf{k}'$ and $z$ axis is $\theta'$.
Here the magnetic field $\mathbf{B}'$ is confined within $xy$ plane.
Let the azimuthal angle of $\mathbf{B}'$ be $\eta'$, and we obtain the relations
between the components of $\mathbf{B}'$ in the systems $123$ and $xyz$.
\begin{equation}
\begin{array}{l}
\sin\theta'_B \sin\phi'_B = \cos\theta' \cos\eta', \\
\sin\theta'_B \cos\phi'_B = \sin\eta', \\
\cos\theta'_B = \sin\theta' \cos\eta'.
\end{array}
\end{equation} 
Then we obtain 
\begin{equation}
\begin{array}{l}
\sin\theta'_B = \left[ 1- \frac{4\gamma^2\theta^2}{(1+\gamma^2\theta^2)^2}
\cos^2\eta' \right]^{1/2}, \\
\cos(2\phi'_B) = \frac{1}{\sin^2\theta'_B}
\left[ \sin^2\eta' -
\left(\frac{1-\gamma^2\theta^2}{1+\gamma^2\theta^2} \right)^2
 \cos^2\eta' \right].
\end{array}
\end{equation}
To obtain the polarization degree of synchrotron radiation from the
random field, we average the Stokes parameters with respect to $\eta'$.
This leads to $\langle j'^U_{\nu'} \rangle = 0$.
Then we can calculate the polarization 
degree by $\Pi_0 = \langle j'^Q_{\nu'} \rangle / \langle j'^I_{\nu'} \rangle
= \Pi_0^{\rm syn} \langle (\sin\theta'_B)^{\alpha+1} \cos(2\phi'_B)\rangle
/\langle (\sin\theta'_B)^{\alpha+1}\rangle$, and the polarization vector
is along axis 1, i.e., the direction perpendicular to $\mathbf{k}'$ and
within the plane including $\mathbf{k}'$ and $\mathbf{\beta}$.


\begin{thebibliography}{}
%
\bibitem[Amati et al.(2002)]{amati02}
Amati,~L. 2002, A\&A, 390, 81
%
\bibitem[Band et al.(1993)]{band93}
Band,~D.~L., et al. 1993, ApJ, 413, 281
%
\bibitem[Begelman \& Sikora(1987)]{begelman87}
Begelman,~M.~C., \& Sikora,~M. 1987, ApJ, 322, 650
%
\bibitem[Black et al.(2007)]{black07}
Black,~J.~K., et al. 2007, NIMA, 581, 755
%
\bibitem[Bloser et al.(2006)]{bloser06}
Bloser,~P.~F., et al. 2006, ChJAS, 6, 393
%
\bibitem[Bloser et al.(2008)]{bloser08}
Bloser,~P.~F., et al. 2008, accepted in NIMA (arXiv:0812.0782)
%
\bibitem[Boggs et al.(2006)]{boggs06}
Boggs,~S.~E., et al. 2006, arXiv:astro-ph/0608532
%
\bibitem[Coburn \& Boggs(2003)]{coburn03}
Coburn~W., \& Boggs,~S.~E. 2003, Nature, 423, 415
%
\bibitem[Costa et al.(2007)]{costa07}
Costa,~E., et al. 2007, in Proc. of SPIE Vol.6686, 30
%
\bibitem[Covino et al.(2004)]{covino04}
Covino,~S., Ghisellini,~G., Lazzati,~D., \& Malesani,~D. 2004,
in ASP Conf. Ser. 312, 169
%
\bibitem[Dai \& Zhang(2005)]{dai05}
Dai,~X., \& Zhang,~B. 2005, ApJ, 621, 875
%
\bibitem[Donaghy(2006)]{donaghy06}
Donaghy,~T.~Q. 2006, ApJ, 645, 436
%
\bibitem[Eichler \& Levinson(2003)]{eichler03}
Eichler,~D., \& Levinson,~A. 2003, ApJ, 596, L147
%
\bibitem[Fendt \& Ouyed(2004)]{fendt04}
Fendt,~C., \& Ouyed,~R. 2004, ApJ, 608, 378
%
\bibitem[Frail et al.(2001)]{frail01}
Frail,~D.~A., et al. 2001, ApJ, 562, L55
%
\bibitem[Genet et al.(2007)]{genet07}
Genet,~F., Daigne,~F., \& Mochkovitch,~R. 2007, MNRAS, 381, 732
%
\bibitem[Ghisellini \& Lazzati(1999)]{ghisellini99}
Ghisellini,~G., \& Lazzati,~D. 1999, MNRAS, 309, L7
%
\bibitem[Ghisellini et al.(2007)]{ghisellini07}
Ghisellini,~G., Ghirlanda,~G., Nava,~L., \& Firmani,~C. 2007, ApJ, 658, L75
%
\bibitem[Granot et al.(1999)]{granot99}
Granot,~J., Piran~T., \& Sari,~R. 1999, ApJ, 513, 679
%
\bibitem[Granot(2003)]{granot03}
Granot,~J. 2003, ApJ, 596, L17
%
\bibitem[Granot \& Taylor(2005)]{granot05}
Granot,~J., \& Taylor,~G.~B. 2005, ApJ, 625, 263
%
\bibitem[Greiner et al.(2008)]{greiner08}
Greiner,~J., et al. 2008, in 
AIP Conference Proceedings, Volume 1000, 620.
%
\bibitem[Gruzinov \& Waxman(1999)]{gruzinov99}
Gruzinov,~A., \& Waxman,~E. 1999, ApJ, 511, 852
%
\bibitem[Hill et al.(2008)]{hill08}
Hill,~J.~E., et al. 2008, in Proc. of 2008 Nanjing GRB Conference
(arXiv: 0810.2499)
%
\bibitem[Hjorth et al.(2003)]{hjorth03}
Hjorth,~J. 2003, Nature, 423, 847
%
\bibitem[Ioka \& Nakamura(2001)]{ioka01}
Ioka,~K., \& Nakamura,~T. 2001, ApJ, 554, L163
%
\bibitem[Ioka \& Nakamura(2002)]{ioka02}
Ioka,~K., \& Nakamura,~T. 2002, ApJ, 570, L21
%
\bibitem[Ioka et al.(2007)]{ioka07}
Ioka,~K., et al. 2007, ApJ, 670, L77
%
\bibitem[Jahoda et al.(2007)]{jahoda07}
Jahoda,~K., et al. 2007, arXiv:astro-ph/0701090
%
\bibitem[Jason et al.(2005)]{jason05}
Jason,~L., et al. 2005, in Proc. of SPIE Vol.5898, 413
%
\bibitem[Kalemci et al.(2007)]{kalemci07}
Kalemci,~E., et al. 2007, ApJS, 169, 75
%
\bibitem[Lazzati et al.(2000)]{lazzati00}
Lazzati,~D., Ghisellini,~G., Celotti,~A., \& Rees,~M.~J. 2000, ApJ, 529, L17
%
\bibitem[Lazzati et al.(2004)]{lazzati04}
Lazzati,~D., Rossi,~E., Ghisellini,~G., Rees,~M.~J. 2004, MNRAS, 347, L1
%
\bibitem[Lazzati(2006)]{lazzati06}
Lazzati,~D. 2006, New Journal of Physics, 8, 131
%
\bibitem[Levinson \& Eichler(2004)]{levinson04}
Levinson,~A., \& Eichler,~D. 2004, ApJ, 613, 1079
%
\bibitem[Lithwick \& Sari(2001)]{lithwick01}
Lithwick,~Y., \& Sari,~R. 2001, ApJ, 555, 540
%
\bibitem[Lyutikov et al.(2003)]{lyutikov03}
Lyutikov,~M., Pariev,~V.~I., \& Blandford,~R.~D. 2003, ApJ, 597, 998
%
\bibitem[Matsumiya \& Ioka(2003)]{matsumiya03}
Matsumiya,~M., \& Ioka,~K. 2003, ApJ, 595, L25
%
\bibitem[McGlynn et al.(2007)]{mcglynn07}
McGlynn,~S., et al. 2007, A\&A, 466, 895
%
\bibitem[Medvedev \& Loeb(1999)]{medvedev99}
Medvedev,~M.~V., \& Loeb,~A. 1999, ApJ, 526, 697
%
\bibitem[M\'{e}sz\'{a}ros(2006)]{meszaros06}
M\'{e}sz\'{a}ros,~P. 2006, Rep.~Prog.~Phys., 69, 2259
%
\bibitem[Mizuno et al.(2005)]{mizuno05}
Mizuno,~T., et al. 2005, Nucl. Instrum. Methods. Phys. Res. Sec. A, 540, 158
%
\bibitem[Mundell et al.(2007)]{mundell07}
Mundell,~C.~G., et al. 2007, Science, 315, 1822
%
\bibitem[Nakar et al.(2003)]{nakar03}
Nakar,~E., Piran,~T., \& Waxman,~E. 2003, JCAP, 10, 5
%
\bibitem[Panaitescu \& Meszaros(2000)]{panaitescu00}
Panaitescu,~A., \& M\'{e}sz\'{a}ros,~P. 2000, ApJ, 544, L17
%
\bibitem[Piran(2005)]{piran05}
Piran,~T. 2005, Rev. Mod. Phys., 76, 1143
%
\bibitem[Porciani \& Madau(2001)]{porciani01}
Porciani,~C., \& Madau,~P. 2001, ApJ, 548, 522
%
\bibitem[Preece et al.(2000)]{preece00}
Preece,~R.~D., et al. 2000, ApJS, 126, 19
%
\bibitem[Produit et al.(2005)]{produit05}
Produit,~N., et al. 2005, Nucl. Instrum. Methods. Phys. Res. Sec. A, 550, 616
%
\bibitem[Racusin et al.(2008)]{racusin08}
Racusin,~J.~L., et al. 2008, Nature, 455, 183
%
\bibitem[Rees \& M\'{e}sz\'{a}ros(1994)]{rees94}
Rees,~M.~J., \& M\'{e}sz\'{a}ros,~P. 1994, ApJ, 430, L93
%
\bibitem[Rutledge \& Fox(2004)]{rutledge04}
Rutledge,~R.~E., \& Fox,~D.~B. 2004, MNRAS, 350, 1288
%
\bibitem[Rybicki \& Lightman(1979)]{rybicki79}
Rybicki,~G.~B., \& Lightman,~A.~P. 1979, Radiative Processes in Astrophysics
(New York: Wiley Interscience)
%
\bibitem[Ryde et al.(2006)]{ryde06}
Ryde,~F., et al. 2006, ApJ, 652, 1400
%
\bibitem[Sagiv et al.(2004)]{sagiv04}
Sagiv,~A., Waxman,~E., \& Loeb,~A. 2004, ApJ, 615, 366
%
\bibitem[Sakamoto et al.(2005)]{sakamoto05}
Sakamoto,~T., et al. 2005, ApJ, 629, 311
%
\bibitem[Sari(1999)]{sari99}
Sari,~R. 1999, ApJ, 524, L43
%
\bibitem[Schady et al.(2007)]{schady07}
Schady,~P., et al. 2007, MNRAS, 380, 1041
%
\bibitem[Shaviv \& Dar(1995)]{shaviv95}
Shaviv,~N.~J., \& Dar,~A. 1995, ApJ, 447, 863
%
\bibitem[Spruit et al.(2001)]{spruit01}
Spruit,~H.~C., Daigne,~F., \& Drenkhahn,~G. 2001, A\&A, 369, 694
%
\bibitem[Stanek et al.(2003)]{stanek03}
Stanek,~K.~Z., et al. 2003, ApJ, 591, L17
%
\bibitem[Taylor et al.(2004)]{taylor04}
Taylor,~G.~B., Frail,~D.~A., Berger,~E., \& Kulkarni,~S.~R. 2004,
ApJ, 609, L1
%
\bibitem[Taylor et al.(2005)]{taylor05}
Taylor,~G.~B., Momjian,~E., Philstrom,~Y., Ghosh,~T., \& Salter,~C.
2005, ApJ, 622, 986
%
\bibitem[Thompson et al.(2007)]{thompson07}
Thompson,~C., M\'{e}sz\'{a}ros,~P., \& Rees,~M.~J. 2007, ApJ, 666, 1012
%
\bibitem[Toma et al.(2005)]{toma05}
Toma,~K., Yamazaki,~R., Nakamura,~T. 2005, ApJ, 635, 481
%
\bibitem[Toma et al.(2008)]{toma08}
Toma,~K., Ioka,~K., \& Nakamura,~T. 2008, ApJ, 673, L123
%
\bibitem[Uhm \& Beloborodov(2007)]{uhm07}
Uhm,~Z.~L., \& Beloborodov,~A.~M. 2007, ApJ, 665, L93
%
\bibitem[Waxman(2003)]{waxman03}
Waxman,~E. 2003, Nature, 423, 388
%
\bibitem[Wigger et al.(2004)]{wigger04}
Wigger,~C., et al. 2004, ApJ, 613, 1088
%
\bibitem[Willis et al.(2005)]{willis05}
Willis,~D.~R., et al. 2005, A\&A, 439, 245
%
\bibitem[Woods \& Loeb(1999)]{woods99}
Woods,~E., \& Loeb,~A. 1999, ApJ, 523, 187
%
\bibitem[Yamazaki et al.(2004)]{yamazaki04}
Yamazaki,~R., Ioka,~K., \& Nakamura,~T. 2004, ApJ, 606, L33
%
\bibitem[Yonetoku et al.(2005)]{yonetoku05}
Yonetoku,~D., Yamazaki,~R., Nakamura,~T., \& Murakami,~T. 2005, 
MNRAS, 362, 1114
%
\bibitem[Zeh et al.(2006)]{zeh06}
Zeh,~A., Klose,~S., \& Kann,~D.~A. 2006, ApJ, 637, 889
%
\bibitem[Zhang \& M\'{e}sz\'{a}ros(2002)]{zhang02}
Zhang,~B., \& M\'{e}sz\'{a}ros,~P. 2002, ApJ, 581, 1236
%
\bibitem[Zhang et al.(2004)]{zhang04}
Zhang,~B., Dai,~X., Lloyd-Ronning,~N.~M., \& M\'{e}sz\'{a}ros,~P.
2004, ApJ, 601, L119
%
\bibitem[Zhang(2007)]{zhang07}
Zhang,~B. 2007, Chin.~J.~Astron.~Astrophys., 7, 1
%
\end{thebibliography}
\end{document}